\documentstyle[eqsecnum,aps,epsfig]{revtex}
\begin{document}
\draft
\title{
Evolution of circular, non-equatorial orbits of Kerr black holes
due to gravitational-wave emission: II. Inspiral trajectories and
gravitational waveforms
}
\author{Scott A.\ Hughes}
\address{
Institute for Theoretical Physics, University of California,
Santa Barbara, CA 93103\\
Theoretical Astrophysics, California Institute of Technology,
Pasadena, CA 91125\\
}
\maketitle
\begin{abstract}
The inspiral of a ``small'' ($\mu \sim 1-100\,M_\odot$) compact body
into a ``large'' ($M \sim 10^{5-7}\,M_\odot$) black hole is a key
source of gravitational radiation for the space-based
gravitational-wave observatory LISA.  The waves from such inspirals
will probe the extreme strong-field nature of the Kerr metric.  In
this paper, I investigate the properties of a restricted family of
such inspirals (the inspiral of circular, inclined orbits) with an eye
toward understanding observable properties of the gravitational waves
that they generate.  Using results previously presented to calculate
the effects of radiation reaction, I assemble the inspiral
trajectories (assuming that radiation reacts adiabatically, so that
over short timescales the trajectory is approximately geodesic) and
calculate the wave generated as the compact body spirals in.  I do
this analysis for several black hole spins, sampling a range that
should be indicative of what spins we will encounter in nature.  The
spin has a very strong impact on the waveform.  In particular, when
the hole rotates very rapidly, tidal coupling between the inspiraling
body and the event horizon has a very strong influence on the inspiral
time scale, which in turn has a big impact on the gravitational wave
phasing.  The gravitational waves themselves are very usefully
described as ``multi-voice chirps'': the wave is a sum of ``voices'',
each corresponding to a different harmonic of the fundamental orbital
frequencies.  Each voice has a rather simple phase evolution.
Searching for extreme mass ratio inspirals voice-by-voice may be more
effective than searching for the summed waveform all at once.
\end{abstract}

\section{Introduction}
\label{sec:intro}

One of the goals of the space-based gravitational-wave detector LISA
{\cite{LISA}} is to make measurements that will probe gravity in the
very strong field.  The most stringent probes will involve binary
black hole systems.  Mergers of comparable mass black holes will be
detectable throughout most of the universe {\cite{flanhughes}}; their
measurements will probe the violent dynamics of two black holes
combining to form a single hole.  LISA will measure hundreds to
thousands of wave cycles in such events, probing their waveforms with
moderate precision.  When the mass ratio of the system is extreme
($M_1/M_2 \simeq 10^{-3} - 10^{-7}$), measurement of the waves probes
the quiescent structure of the Kerr black hole spacetime.  LISA should
measure tens of thousands to millions of cycles from such inspirals
(depending mostly on mass ratio), and thus may be able to measure
their waves --- and probe black hole spacetimes --- with extremely
high precision.  Extreme mass ratio inspirals will be the focus of
this analysis.

The case for studying extreme mass ratio inspirals has been argued at
length elsewhere; I refer the reader to the Introductions of Refs.\
{\cite{flanhughes}} and {\cite{finnthorne}} for detailed discussion.
Briefly, detailed studies {\cite{sigurdssonrees,sigurdsson}} have
shown that compact stellar remnants in the central cusp of galaxies
are scattered into very tight, eccentric orbits of the galaxies' black
hole at a rate of at least several events per year out to a Gigaparsec
(perhaps as high as an event per month).  The final year or so of
inspiral (depending on the large black hole's mass and spin, and on
the binary's mass ratio) will occur as the small body spirals through
the most extreme strong-field region of the black hole spacetime, from
several horizon radii down to orbits just outside the hole's event
horizon.

By tracking the phase evolution of the waves over the many cycles
radiated in this year, it should be possible to measure strong-field
characteristics of the black hole spacetime.  Fintan Ryan
{\cite{ryan_meas}} has shown that these waves can be used to
reconstruct the multipolar structure of the black hole.  By the ``no
hair'' theorem, a Kerr black hole's multipole moments are uniquely
determined by its mass $M$ and spin $a$.  Measuring more than two
moments tests whether the massive object at the core of a galaxy is a
black hole, or whether it is some more exotic compact object, such as
a soliton star.

Analyzing how well one can measure the multipoles of a massive compact
object in this manner means computing gravitational waves from the
inspiral of realistic small bodies on inclined, eccentric trajectories
about bodies with arbitrary multipole moments.  We are rather far from
being able to analyze such a system.  For now, I require that the
massive body be a black hole --- this greatly simplifies the
description of its orbits.  I also take the inspiraling body to be
pointlike and nonspinning, sidestepping issues of spin-spin and
spin-orbit coupling which complicate the inspiral and may in fact lead
to chaos {\cite{janna,hartlphinney}}.  Finally, because rigorous
strong-field radiation reaction is currently being developed
{\cite{forcepapers}}, I use a ``poor man's radiation reaction''
approach, balancing the flux of energy and angular momentum in
gravitational waves with the change in orbital energy and angular
momentum.  This formalism works well if the inspiral is adiabatic: the
change in orbital parameters due to wave emission must be slow.  (A
more precise definition of adiabatic and ``slow'' is given in Sec.\
{\ref{subsec:circ_radreact}}.)  At present, the flux-balance approach
cannot be used to find the change in the third constant of Kerr orbits
(the Carter constant $Q$), so I require that the orbit be initially
``circular'' (of constant Boyer-Lindquist radius).  These orbits do
not become eccentric under adiabatic radiation reaction
{\cite{ryan_circ,dan_amos,mino_thesis}}, constraining the system
enough that $Q$'s evolution can be inferred from the changes in the
orbital energy and angular momentum alone.

Given the large number of restrictions imposed, it is clear this
analysis is rather far from the ultimate goal of laying the
foundations for compact body multipole measurement with LISA.
However, it is sufficient to begin developing gravitational waveforms
and probing the inspiral trajectories for non-trivial orbits.  These
orbits are colored rather strongly by the orbital frequencies and
their harmonics, and, despite the numerous restrictions, have a fairly
ornate structure.  This analysis is a useful way to produce
gravitational waveforms that have a character similar to what we are
likely to observe with LISA.  They are restricted enough that they can
be calculated with tools available now, but have enough structure that
they can be used for developing data analysis tools and understanding
how measurement of extreme mass ratio inspirals with LISA will work in
practice.

In the remainder of the Introduction, I summarize the structure and
results of the paper.  I first review the properties of circular,
inclined orbits of Kerr black holes and the radiation reaction
formalism used here.  This material is presented in far greater detail
in Refs.\ {\cite{paperI}} and {\cite{nhs}}.  This paper just presents
equations and points the reader to the relevant literature for a
detailed derivation.  In the end, radiation reaction data are
presented as vectors $(\dot r,\dot\iota)$ living at orbital
coordinates $(r,\iota)$.  These vectors describe how radiation drives
the body to spiral through a sequence of orbits into the hole.  The
wave itself is developed from a set of complex amplitudes $Z^H_{lmk}$
which constitute a multipole expansion of the gravitational-wave
strain measured by distant observers.  This fully sets up this
analysis.

Inspiral trajectories are constructed by beginning at some coordinate
$(r_0,\iota_0)$ and stepping inward along the data $(\dot
r,\dot\iota)$ until the body hits the last dynamically stable orbit
and plunges into the hole.  The data are actually computed on a grid
that covers the parameter space of allowed orbits deep in the strong
field, and are interpolated to get radiative effects off the grid.
Section {\ref{subsec:grid}} first describes how the data grid is set
up, as well as which data are stored on it.  Section
{\ref{subsec:integrate}} then briefly describes how the integration is
done.

\begin{figure}[ht]
\begin{center}
\epsfig{file = 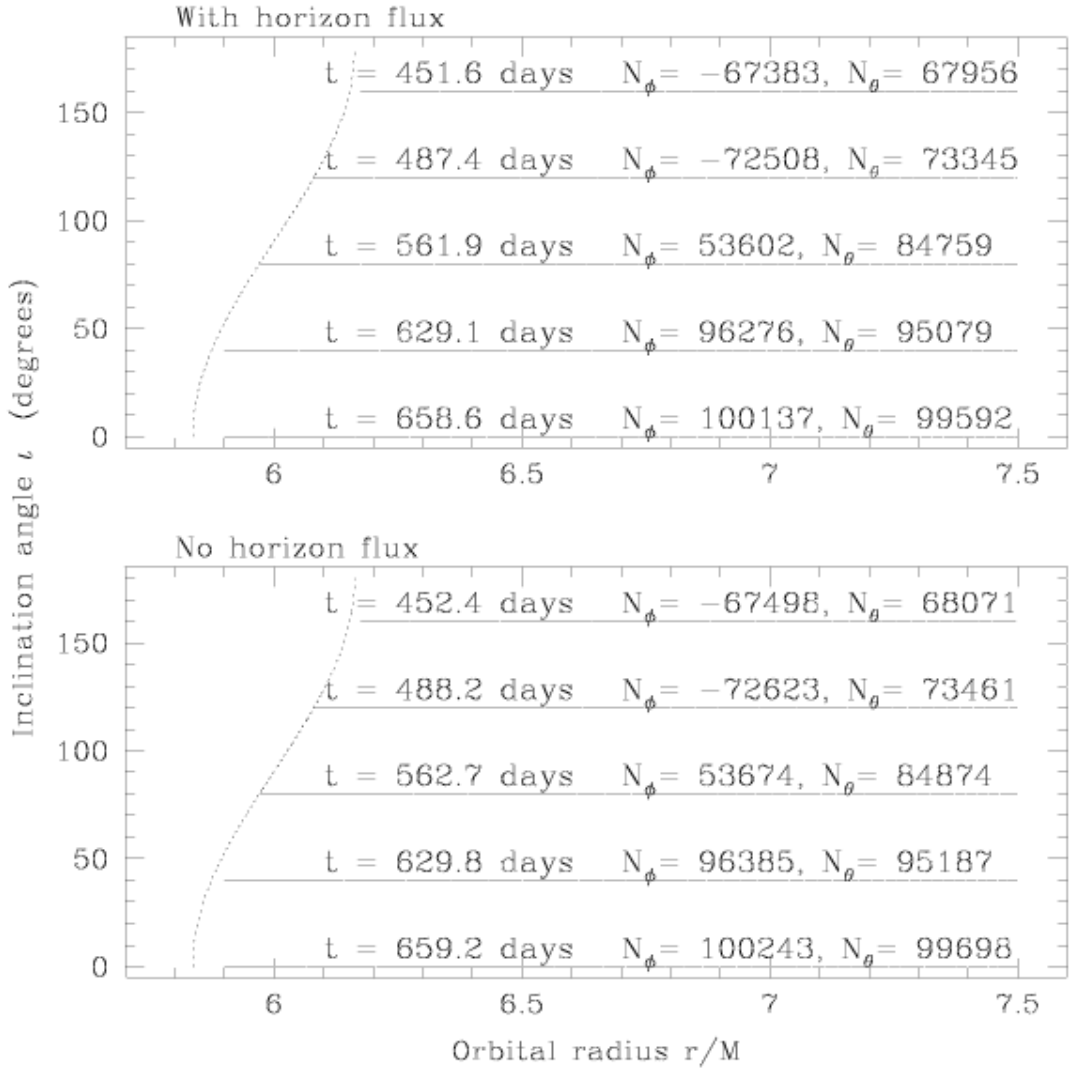, width = 15cm}
\caption{\label{fig:traj_0.05}
Inspiral trajectories in the strong field of a Kerr black hole with $a
= 0.05M$.  The top panel includes the effects of down-horizon flux;
the bottom panel does not.  In this case, inspiral is quicker when the
horizon flux is included: the hole's event horizon acts as an energy
sink, as simple intuition suggests it should.  This is because the
tidal bulge raised on the hole by the orbiting body tends to lag,
rather than lead, the orbit.  Thus the bulge's torque on the orbit
opposes the orbital motion, causing it to spiral in more quickly.
}
\end{center}
\vskip -0.5cm
\end{figure}

Results are given in Secs.\ {\ref{sec:trajectories}} and
{\ref{sec:waveforms}}.  First I discuss the inspiral trajectories in
Sec.\ {\ref{sec:trajectories}}; they and their various features are
displayed for several choices of black hole spin in Figs.\
{\ref{fig:traj_H_0.998}}, {\ref{fig:traj_nH_0.998}},
{\ref{fig:traj_0.3594}}, {\ref{fig:traj_0.8}}, and
{\ref{fig:traj_0.05}}.  During inspiral lasting 1--2 years, the
inspiraling body executes several hundred thousand orbits in the
extreme strong field of the massive black hole.  This is not a
surprise, and has of course been known for quite some time.  More
interestingly, the inclination angle barely changes during inspiral,
particularly when spin $a \alt 0.8 M$.  This gives further credence to
a suggestion by Curt Cutler that it would be worthwhile to explore
``faking'' the inspiral of generic Kerr orbits by holding the
inclination angle fixed --- this would fix the change in the Carter
constant provided the changes in $E$ and $L_z$ were known.

The influence of the black hole's event horizon on the inspiral is
very strong.  A very useful way to understand this effect is that the
inspiraling body tidally distorts the black hole, causing the horizon
to bulge.  This bulge then exerts a torque back on the body.  This
viewpoint is explained at a very pedagogically accessible level in
Chapter VII of Ref.\ {\cite{membrane}}, which in turn is based on the
work of Hawking and Hartle {\cite{hawkinghartle}} and Hartle
{\cite{hartle}}.  If the hole is rapidly rotating, this torque tends
to significantly slow the inspiral, transferring some of the hole's
rotational kinetic energy to the orbital motion.  Inspiral can be
prolonged for several weeks ($\sim 5\%$ of the total inspiral time) by
this effect, adding tens of thousands of additional orbits.  By
contrast, if the hole is rotating slowly, the torque exerted by the
horizon bulge speeds up the inspiral.  This fascinating coupling of
the horizon to the inspiraling body's dynamics is an example of how
strong-field features of the Kerr spacetime can be seen in
gravitational waves.

Section {\ref{sec:waveforms}} discusses the waves produced as the body
spirals inward.  It should be strongly emphasized that computing
waveforms is more difficult than computing inspiral trajectories.  The
waveforms require both the magnitude and phase of the complex
amplitudes $Z^H_{lmk}$ to high precision; the trajectories require
just the magnitude.  As such, the results shown in Sec.\
{\ref{sec:waveforms}} should be taken as an important first step in
generating extreme mass ratio inspiral waves, but can --- and should!
--- be improved dramatically.

Multiple harmonics of the orbital frequencies $\Omega_\phi$ and
$\Omega_\theta$ are very important; cf.\ Fig.\ {\ref{fig:waveform}}.
Following a suggestion of Sam Finn, I write the waveform as a sum over
many ``voices'': $h(t) = \sum h_{lmk}(t)$.  Each voice $h_{lmk}(t)$
corresponds to a particular harmonic of the orbital frequencies, and
has its own amplitude and phase evolution.  Writing the waveform in
this manner emphasizes that the phase evolution of {\it each voice} is
rather simple, though the sum may be complicated and difficult to
follow.  It is likely this multi-voice structure will carry over to
the generic case, adding an additional index for harmonics of the
radial frequency $\Omega_r$.  The analysis of LISA data may be
facilitated by searching for extreme mass ratio inspirals
voice-by-voice, rather than searching for the ``chorus'' of all voices
at the same time.  Concluding discussion, including suggestions for
future work and extensions to this analysis, is given in Sec.\
{\ref{sec:conclusion}}.

Throughout this paper, an overdot denotes $d/dt$, and a prime denotes
$\partial/\partial r$.  An overbar indicates complex conjugation.  The
quantities $t$, $r$, $\theta$, and $\phi$ refer to the Boyer-Lindquist
coordinates.  I use units with $G = 1 = c$ throughout.

\section{Review: Circular orbits and radiation reaction}
\label{sec:circ_motion}

\subsection{Geodesic orbits}
\label{subsec:circ_geod}

Orbits of Kerr black holes are specified by choosing their energy $E$,
axial angular momentum $L_z$, and Carter constant $Q$.  The values
used here have been divided by the small body's mass $\mu$ ($E$ and
$L_z$) or $\mu^2$ ($Q$) and are thus the specific energy, angular
momentum and Carter constant.  Having chosen these constants (and
initial conditions), the orbit is then governed by geodesic equations
for the small body's Boyer-Lindquist coordinates ($t$, $r$, $\theta$,
$\phi$) as a function of proper time $\tau$ measured along its
worldline; see {\cite{mtw}}.

To describe circular orbits, it is useful to introduce the function
$R \equiv \Sigma^2 (dr/d\tau)^2$:
\begin{equation}
R = \left[E(r^2 + a^2) - a L_z\right]^2 -
\Delta\left[r^2 + (L_z - a E)^2 + Q\right]\;.
\label{eq:def_of_R}
\end{equation}
(The function $\Sigma = r^2 + a^2\cos^2\theta$, and $\Delta = r^2 - 2
M r + a^2$.)  Circular orbits satisfy $R = 0$ and $R' = 0$; stable
circular orbits satisfy in addition $R'' \le 0$.  Given the radius $r$
and one of the three constants $E$, $L_z$, or $Q$, it is
straightforward to solve the system $R = 0$, $R' = 0$ for the other
two.  In this paper, I choose $r$ and $E$ and then solve for $L_z$ and
$Q$:
\begin{eqnarray}
L_z(r,E) &=& {E M(r^2 - a^2) - \Delta\sqrt{r^2(E^2 - 1) + r M}\over
a(r - M)}\;,
\label{eq:Lz_of_E}\\
Q(r,E) &=& {\left[(a^2 + r^2)E - a
L_z(r,E)\right]\over\Delta} - \left[r^2 + a^2 E^2 - 2 a E L_z(r,E) +
L_z(r,E)^2\right]\;.
\label{eq:Q_of_E}
\end{eqnarray}
[Note that in Ref.\ {\cite{nhs}}, the $a^2 E^2$ inside the second set
of square brackets is incorrectly written $a^2 E$.]

At a given radius, two orbits bound the behavior of all stable
circular orbits.  The most-bound orbit is the prograde equatorial
orbit ($\iota = 0$).  Its constants are {\cite{bpt}}
\begin{eqnarray}
E^{\rm mb} &=& {{1 - 2 v^2 + q v^3}\over\sqrt{1 - 3v^2 + 2q v^3}}\;,
\label{eq:Emb}\\
L_z^{\rm mb} &=& rv {{1 - 2 q v^3 + q^2 v^4}\over\sqrt{1 - 3v^2 + 2q v^3}}\;,
\label{eq:Lzmb}\\
Q^{\rm mb} &=& 0\;,
\label{eq:Qmb}
\end{eqnarray}
where $v \equiv \sqrt{M/r}$ and $q \equiv a/M$.  Prograde equatorial
orbits exist at all radii outside $r_{\rm pro}$, where
\begin{eqnarray}
r_{\rm pro}/M &=& 3 + Z_2 + \left[(3 - Z_1)(3 + Z_1 + 2
Z_2)\right]^{1/2}\;,
\label{eq:r_pro}\\
Z_1 &=& 1 + \left[1 - (a/M)^2\right]^{1/3}\left[(1 + a/M)^{1/3} +
(1 - a/M)^{1/3}\right]\;,
\label{eq:Z1}\\
Z_2 &=& \left[3(a/M)^2 + Z_1^2\right]^{1/2}\;.
\label{eq:Z2}
\end{eqnarray}
This is also the radius of the LSO when $\iota = 0^\circ$ --- no
stable circular orbits exist inside $r_{\rm pro}$.  If $r < r_{\rm
ret}$, where
\begin{equation}
r_{\rm ret}/M = 3 + Z_2 + \left[(3 - Z_1)(3 + Z_1 + 2
Z_2)\right]^{1/2}\;,
\label{eq:r_ret}
\end{equation}
then the least-bound orbit is on the LSO, and the constants $(E^{\rm
lb}, L_z^{\rm lb}, Q^{\rm lb})$ are found by numerically solving the
system of equations $R = 0$, $R' = 0$, $R'' = 0$.  If $r > r_{\rm
ret}$, the least-bound orbit is the retrograde equatorial orbit
($\iota = 180^\circ$), and its constants are
\begin{eqnarray}
E^{\rm lb} &=& {{1 - 2 v^2 - q v^3}\over\sqrt{1 - 3v^2 - 2q v^3}}\;,
\label{eq:Elb}\\
L_z^{\rm lb} &=& -rv{{1 + 2 q v^3 + q^2 v^4}\over\sqrt{1 - 3v^2 - 2q v^3}}\;,
\label{eq:Lzlb}\\
Q^{\rm lb} &=& 0\;.
\label{eq:Qlb}
\end{eqnarray}
Orbits with $Q \ne 0$ are inclined with respect to the equatorial
plane; a useful definition of the inclination angle $\iota$
is\footnote{As discussed in Ref.\ {\cite{paperI}}, this angle does not
necessarily accord with intuitive notions of inclination angle.  For
example, except when $a = 0$, $\iota$ is {\it not} the angle at which
most observers would see the small body cross the equatorial plane.}
\begin{equation}
\cos\iota = {L_z\over\sqrt{L_z^2 + Q}}\;.
\label{eq:cosiota}
\end{equation}

Mapping out all stable circular orbits at some radius is fairly
straightforward.  First, calculate the constants that describe the
most-bound orbit [Eqs.\ (\ref{eq:Emb})--(\ref{eq:Qmb})].  Second,
calculate the constants for the least-bound orbit [solving the system
$R = 0$, $R' = 0$, $R'' = 0$ if $r < r_{\rm ret}$; using Eqs.\
(\ref{eq:Elb})--(\ref{eq:Qlb}) otherwise].  Finally, vary $E$ from its
most-bound to its least-bound extreme.

Circular orbits are periodic, with two (generally incommensurate)
frequencies $\Omega_\theta$ and $\Omega_\phi$, related to the small
body's motion in the $\theta$ and $\phi$ coordinates.  These two
frequencies (and their harmonics) strongly stamp the gravitational
waveform.  Typically, $\Omega_\phi > \Omega_\theta$; the differences
are due to the oblate geometry of a rotating black hole and frame
dragging (which can greatly augment $\Omega_\phi$).  A detailed
derivation of these frequencies is given in Sec.\ IIC of Ref.\
{\cite{paperI}}.

\subsection{Radiative corrections}
\label{subsec:circ_radreact}

The radiation reaction formalism used here is based on the Teukolsky
equation {\cite{teuk73}}, which governs the evolution of the complex
Weyl curvature scalar related to radiative perturbations, $\psi_4$.
This section briefly outlines how the radiative corrections are
computed, particularly the quantities relevant to this analysis.
Reference {\cite{paperI}} discusses this material in greater detail.

The ``master equation'' for the evolution of $\psi_4$ is separated
with the multipolar decomposition {\cite{teuk73}}
\begin{equation}
\psi_4 = {1\over(r - i a\cos\theta)^4}\int_{-\infty}^\infty d\omega
\sum_{lm} R_{lm\omega}(r)\, _{-2}S^{a\omega}_{lm}(\theta)e^{i(m\phi
- \omega t)}\;.
\label{eq:psi4decomp}
\end{equation}
The $t$ and $\phi$ dependences are trivial.  The function
$_{-2}S^{a\omega}_{lm}(\theta)$ is a spin-weighted spheroidal
harmonic.  It is useful for describing the $\theta$ dependence of a
radiation field with spin weight $-2$ in an oblate geometry.  An
effective algorithm for calculating this function is given in Appendix
A of Ref.\ {\cite{paperI}}.

Computing the radial function $R_{lm\omega}$ takes some effort.  This
function obeys the Teukolsky equation:
\begin{equation}
\Delta^2{d\over dr}\left({1\over\Delta}{dR_{lm\omega}\over dr}\right)
- V(r) R_{lm\omega} = -{\cal T}_{lm\omega}(r)\;.
\label{eq:teuk}
\end{equation}
This equation is in self-adjoint form, and so can be solved with
Green's functions {\cite{arfken}}.  The homogeneous Teukolsky equation
has two independent solutions, $R^H_{lm\omega}(r)$ (which obeys the
boundary condition that radiation must be purely ingoing at the event
horizon) and $R^\infty_{lm\omega}(r)$ (which obeys the condition that
radiation must be purely outgoing at infinity).  From these solutions
and from the source function ${\cal T}_{lm\omega}(r)$, one can write
down the general solution $R_{lm\omega}(r)$:
\begin{equation}
R_{lm\omega}(r) = Z^H_{lm\omega} R^\infty_{lm\omega}(r) +
Z^\infty_{lm\omega} R^H_{lm\omega}(r)\;.
\label{eq:teuk_gensoln}
\end{equation}
A detailed description of how one computes $Z_{lm\omega}^H$,
$Z_{lm\omega}^\infty$, $R_{lm\omega}^H$, and $R_{lm\omega}^\infty$
(and how they relate to the source function ${\cal T}_{lm\omega}$) is
given in Ref.\ {\cite{paperI}}.  The most important details for this
paper are as follows.  First, for circular orbits, the source and all
quantities derived from it are describable as harmonics of
$\Omega_\theta$ and $\Omega_\phi$.  Defining $\omega_{mk} =
m\Omega_\phi + k\Omega_\theta$, the functions $Z_{lm\omega}^H$ and
$Z_{lm\omega}^\infty$ can be decomposed into harmonics:
\begin{equation}
Z^{H,\infty}_{lm\omega} = \sum_k Z^{H,\infty}_{lmk}
\delta(\omega - \omega_{mk})\;.
\label{eq:Zed_decomp}
\end{equation}
Second, as $r \to r_+ = M + \sqrt{M^2 - a^2}$ (the coordinate of the
event horizon), $Z_{lm\omega}^H \to 0$ and
$Z_{lm\omega}^\infty\to\mbox{constant}$; as $r\to \infty$,
$Z_{lm\omega}^\infty\to 0$ and $Z_{lm\omega}^H\to\mbox{constant}$.
The behavior of the radiation at infinity thus depends only on
$Z_{lmk}^H$, and at the horizon only on $Z_{lmk}^{\infty}$.  The
gravitational waveform is
\begin{equation}
h_+ - ih_\times =\sum_{lmk}{Z^H_{lmk}\over\omega_{mk}^2}
{_{-2}S}^{a\omega_{mk}}_{lm}(\vartheta) e^{i(m\varphi -
\omega_{mk}t)}\;,
\label{eq:waveform}
\end{equation}
and the fluxes are
\begin{eqnarray}
\left({dE\over dt}\right)_{\rm rad}^{\infty,H} &=&
\sum_{lmk}\alpha_{lmk}^{\infty,H}
{|Z^{H,\infty}_{lmk}|^2\over4\pi\omega_{mk}^2}\;,
\nonumber\\
\left({dL_z\over dt}\right)_{\rm rad}^{\infty,H} &=&
\sum_{lmk}\alpha_{lmk}^{\infty,H}
{m|Z^{H,\infty}_{lmk}|^2\over4\pi\omega_{mk}^3}\;.
\label{eq:Edot_Lzdot_rad}
\end{eqnarray}
By assumption, the change in the orbital energy and angular momentum
is opposite in sign to that radiated: $\dot E^{\rm orbit} = -\dot
E^{\rm rad}$, ${\dot L}_z^{\rm orbit} = -{\dot L}_z^{\rm rad}$.  Note
the somewhat confusing reversal of $H$ and $\infty$ on the $Z_{lmk}$
coefficients; this follows from the definition in Eq.\
(\ref{eq:teuk_gensoln}).  The factor $\alpha^H_{lmk}$ is a rather
messy coefficient that follows from transforming the Kinnersley null
tetrad (which is used to construct $\psi_4$ {\cite{pressteuk}}) to the
Hawking-Hartle null tetrad {\cite{hawkinghartle}} (which is well
behaved on the event horizon); see Ref.\ {\cite{paperI}} for further
details.  (The factor $\alpha^\infty = 1$; it is introduced so that
these formulas can be written compactly.)  As noted in
{\cite{paperI}}, there is a symmetry relation between quantities at
$(l,m,k)$ and $(l,-m,-k)$:
\begin{equation}
Z^{H,\infty}_{l-m-k} = (-1)^{l+k}{\bar Z}^{H,\infty}_{lmk}\;.
\label{eq:zed_symmetry}
\end{equation}
This relationship can be used to reduce computation time --- compute
half the multipoles, use symmetry to get the other half.
Alternatively, it can be used to improve computational accuracy ---
compute all the multipoles, use symmetry to reduce error:
\begin{equation}
Z^{H,\infty}_{lmk} = {1\over2}\left[Z^{H,\infty}_{lmk;\,{\rm comp}} +
(-1)^{l+k}{\bar Z}^{H,\infty}_{l-m-k;\,{\rm comp}}\right]\;.
\label{eq:zed_symm_applied}
\end{equation}
This second approach was used here.

A large number of terms were included when evaluating Eq.\
(\ref{eq:Edot_Lzdot_rad}).  A detailed discussion of the truncation
criterion used here is given in Sec.\ VA of Ref.\ {\cite{paperI}}; in
the language of that paper, $\epsilon_k = 10^{-4}$ and $\epsilon_l =
10^{-3}$.  The truncation error in $\dot E^{H,\infty}$ and ${\dot
L}_z^{H,\infty}$ is therefore roughly $10^{-3}$.  Obtaining this level
of accuracy required summing to at least $l = 8$ (for innermost orbits
of the $a = 0.998M$ black hole, the sums were taken to $l = 19$).  For
each value of $l$, between 10 and 30 values of $k$ were included.  In
all cases, the sum over $m$ ranged from $-l$ to $l$.

Using the Teukolsky equation to compute $\dot E^{H,\infty}$ and ${\dot
L}_z^{H,\infty}$ requires that the inspiral be adiabatic: over orbital
timescales, the trajectory must be nearly geodesic.  This requirement
enters through the source term, which depends on the inspiraling
body's worldline.  This is potentially the setup for a Catch-22: we
need the worldline in order to compute the body's radiative
corrections, but the corrections determine that worldline!  The
solution is to take the worldline to be geodesic for computing the
source.  We thus use the zeroth order (in $\mu/M$) geodesic motion to
compute the first order radiative corrections.  This is a good
approximation only when the radiative change in any orbital quantity
$\chi$ over an orbit is much less than $\chi$: $\dot\chi T_{\rm orb}
\ll \chi$.  Very useful quantities to monitor are the orbital
frequencies, $\Omega_\phi$ and $\Omega_\theta$.  Because $T_\theta =
2\pi/\Omega_\theta$, and defining $T_\phi = 2\pi/\Omega_\phi$, the
adiabaticity condition for these quantities can be written
\begin{equation}
{\cal N}_{\phi,\theta} \equiv
{1\over2\pi}{\Omega_{\phi,\theta}^2\over{\dot\Omega}_{\phi,\theta}}
\gg 1\;,
\label{eq:calN}
\end{equation}
These parameters are closely related to the number of accumulated
orbits in $\phi$ and $\theta$:
\begin{equation}
N_{\phi,\theta} = {1\over2\pi}\int dt\,\Omega_{\phi,\theta}
= \int d\ln\Omega_{\phi,\theta}\,{\cal N}_{\phi,\theta}\;.
\label{eq:adiabat_orb_rel}
\end{equation}
The condition ${\cal N}_{\phi,\theta} \gg 1$ tells us that the orbits
accumulated by the body as it passes through the orbital frequency
band centered on $\Omega_{\phi,\theta}$ and of width
$\delta\Omega_{\phi,\theta} =\Omega_{\phi,\theta}$ must be very large.
More simply, the body must spend many orbits near any point in its
orbital phase space.

The Teukolsky equation can only tell us about the changes in $E$ and
$L_z$.  To fully describe the small body's spiral in, we impose
circularity: adiabatic radiation reaction changes an orbit's radius
and inclination angle, but it does not make the orbit eccentric
{\cite{ryan_circ,dan_amos,mino_thesis}} --- circular orbits remain
circular.  Imposing this rule gives a relatively simple relationship
between ($\dot Q,\dot r$) and ($\dot E, {\dot L}_z$); see Ref.\
{\cite{paperI}}, Sec.\ IIIA for details.  Once $\dot Q$ and ${\dot
L}_z$ are known, it is simple to get the rate of change of $\iota$
using Eq.\ (\ref{eq:cosiota}); cf.\ Eqs.\ (3.7) and (3.8) of Ref.\
{\cite{paperI}}.  The data $(\dot r, \dot\iota)$ are the foundation
for all the inspiral trajectories discussed in the next section.

Letting $\chi$ stand for any of the quantities ($E,L_z,Q,r$), the
total change in $\chi$ is found by summing the change due to radiation
flux out to infinity and due to radiation flux down the event horizon:
\begin{equation}
\dot\chi = \dot\chi^\infty + \eta \dot\chi^H\;.
\label{eq:turnHflux_onoff}
\end{equation}
The parameter $\eta$ (which clearly should equal 1) is introduced so
that we can ``turn off'' the flux down the horizon.  As we shall see
later, setting $\eta = 0$ is a very interesting test of how strongly
the black hole's event horizon influences inspiral.

\section{Computing inspiral trajectories and waveforms}
\label{sec:computing_traj_wave}

Using the techniques described in Sec.\ {\ref{sec:circ_motion}}, it is
straightforward to compute the inspiral trajectory of a body orbiting
a black hole and the gravitational waveforms generated in that
inspiral.  This is done in two steps.  First, a ``grid'' of radiation
reaction data is built in the phase space of allowed orbits in the
black hole's strong field.  The coordinates in this phase space are
$(r,\iota)$.  The radiation reaction data are the vectors along which
gravitational-wave emission drives the orbit, $(\dot r,\dot\iota)$,
plus the amplitude and frequency of the gravitational waveform,
$Z^H_{lmk}$ and $\omega_{mk}$.  I describe how this grid is built in
Sec.\ {\ref{subsec:grid}}.  Next, the radiation reaction data are
integrated to compute the inspiraling body's trajectory.  The data are
interpolated from discrete orbital coordinates on the grid to
arbitrary points in the orbital phase space using a two dimensional
cubic spline {\cite{numrec}}.  I briefly describe how this is done in
Sec.\ {\ref{subsec:integrate}}.

\subsection{Making a radiation reaction grid}
\label{subsec:grid}

Making a grid of radiation reaction data is for the most part
straightforward.  One discretizes the orbital phase space, choosing
indices $i$ and $j$ such that radiation reaction data live at points
$(r_{ij},\iota_{ij})$.  Then, one uses the techniques summarized in
Sec.\ {\ref{sec:circ_motion}} to compute the data.  I describe here
the algorithm used to discretize the orbital phase space, and also the
actual form of the data (which are somewhat massaged so that they can
be interpolated with as little error as possible).

After some experimentation, I have found it convenient to make the
grid evenly spaced in radius, and evenly spaced {\it at each radius}
in the orbital energy.  The discretized radius is thus
\begin{equation}
r_{ij} \equiv r_j = r_{\rm min} + j\,\delta r\;.
\label{eq:discret_rad}
\end{equation}
The parameter $r_{\rm min}$ is chosen to be as close as is convenient
to $r_{\rm pro}$ [cf.\ Eq.\ (\ref{eq:r_pro})].  This creates a small
gap in data coverage near the LSO at small values of $\iota$.  I have
used $\delta r = 0.1 M$ in all computations; as will be discussed
shortly, more intelligent choices can be made.  The index $j$ can in
principle be made arbitrarily large.  For the results discussed in
Secs.\ {\ref{sec:trajectories}} and {\ref{sec:waveforms}}, I set
$j_{\rm max}\sim 20-30$.  For a $10^6\,M_\odot$ black hole and an
inspiraling body with $\mu = 1\,M_\odot$, this leads to an inspiral
lasting about $630$ days --- typical observation time for a LISA
mission.

The discretized energy is
\begin{equation}
E_{ij} = E_{\rm mb} + (i - 1) \delta E_j\;,
\label{eq:discret_energy}
\end{equation}
where $\delta E_j = \left[E_{\rm lb}(r_j) - E_{\rm
mb}(r_j)\right]/(i_{\rm max} - 1)$.  I have used $i_{\rm max} = 9$.
The parameter space coordinates $(r_j, E_{ij})$ fully determine the
orbit at grid point $(i,j)$.  Using the relations given in Sec.\
{\ref{sec:circ_motion}}, we can remap this coordinate to any
convenient parameterization.  I will typically write the grid points
as $(r_j,\iota_{ij})$, with the understanding that this is a remapping
from the coordinates $(r_j, E_{ij})$ that are directly generated.

A better choice for $\delta r$ can be made by studying how $Z^H_{lmk}$
is computed from solutions of the Teukolsky equation.  From Eq.\ (4.8)
of Ref.\ {\cite{paperI}}, we see that
\begin{equation}
Z^H_{lm\omega} \propto \int_{r_+}^r dr'\,
R^H_{lm\omega}(r'){\cal T}_{lm\omega}(r')/\Delta(r')^2\;.
\label{eq:ZedHdef}
\end{equation}
Here, $R^H_{lm\omega}$ is the solution to the source-free Teukolsky
equation that is purely ingoing at the event horizon, ${\cal
T}_{lm\omega}(r)$ is the Teukolsky source term, and $\Delta(r) = r^2 -
2 M r + a^2$.  The coefficient $Z^H_{lmk}$ is then built from a
harmonic decomposition of $Z^H_{lm\omega}$.  Both ${\cal
T}_{lm\omega}$ and $\Delta$ vary relatively slowly with respect to
$r$.  The function $R^H_{lm\omega}$, on the other hand, oscillates
with a phase factor that is roughly $e^{-ipr^*}$, where
\begin{equation}
r^*(r) = r + {2 M r_+\over r_+ - r_-}\ln{r - r_+\over 2M} - 
{2 M r_-\over r_+ - r_-}\ln{r - r_-\over 2M}\;,
\end{equation}
and where $p = \omega - m\Omega_H$, the mode frequency modified by the
hole's spin frequency $\Omega_H = a/2Mr_+$ [cf.\ Eq.\ (4.4) of Ref.\
{\cite{paperI}}].  The function $r^*(r)$ is the Kerr ``tortoise
coordinate'', which often appears in studies of radiation propagation
near black holes.  Notice that $dr^*/dr$ gets quite large near the
black hole.  This tells us that $e^{-ipr^*}$ begins oscillating very
rapidly as the horizon is approached.  The phase of $Z^H_{lmk}$ is
likely to change by $\sim\pi$ radians over a lengthscale $\delta
r^*\sim {1/p_{mk}}$.  This suggests laying out the grid evenly spaced
in $r^*$, or using a denser grid in $r$, with spacing
\begin{equation}
\delta r \simeq \left(\left|{dr^*\over dr}\right|\right)^{-1}
{1\over|p_{LK}|}\;.
\label{eq:suggested_dr}
\end{equation}
Here $L$ is the maximum $l$ index included in the calculation, and $K$
is the maximum $k$ index.  This would create a grid that is far more
densely sampled than that discussed here, especially for rapidly
spinning holes.  The rather crude choice $\delta r = 0.1 M$ places
strict limits on waveform accuracy; this is an obvious starting point
for improvements to this analysis.

Some care must be taken to put data on the grid that are useful for
generating the inspiral trajectory.  For the radiation reaction
quantities, I have found it useful to store $d\cos\iota/dt$ and
\begin{equation}
\left[\cos\iota -\cos\iota_{\rm LSO}(r)\right]\dot r\equiv\dot\rho\;,
\label{eq:rhodot}
\end{equation}
rather than $(\dot r,\dot\iota)$.  The quantity $d\cos\iota/dt$ is
useful simply because it is more naturally related to the rates of
change ${\dot L}_z$ and $\dot Q$.  The quantity $\dot\rho$ on the
other hand accounts for the fact that $\dot r$ diverges as the LSO is
approached --- the prefactor $\cos\iota - \cos\iota_{\rm LSO}(r)$
nicely clears out this divergence.  This makes it possible for $\dot
r$ to be very accurately interpolated from $\iota = 0^\circ$ to
$\iota_{\rm LSO}(r)$.  For the gravitational waveform, the frequencies
$\Omega_\phi$ and $\Omega_\theta$ are stored, as are the magnitude
${\cal A}_{lmk}$ and phase $\Phi_{lmk}$ of the complex amplitude,
\begin{equation}
Z^H_{lmk} = {\cal A}_{lmk}\exp(i\Phi_{lmk})\;.
\label{eq:amp_phase_def}
\end{equation}
For equatorial orbits, $Z^H_{lmk} = 0$ when $k \ne 0$.  The numerical
code which computes $\Phi_{lmk}$ from $Z^H_{lmk}$ erroneously assigns
$\Phi_{lmk} = 0$ in this case.  This is because the code cannot
evaluate $\Phi_{lmk} = \arctan[{\rm Im}(Z^H_{lmk})/{\rm
Re}(Z^H_{lmk})]$ in the limit $Z^H_{lmk} \to 0$.  To get around this
problem, $\Phi_{l,m,k\ne0}$ is only stored for non-equatorial orbits.
Near the equator, $\Phi_{l,m,k\ne0}$ can then be computed quite well
using extrapolation.

\begin{figure}[ht]
\begin{center}
\epsfig{file = 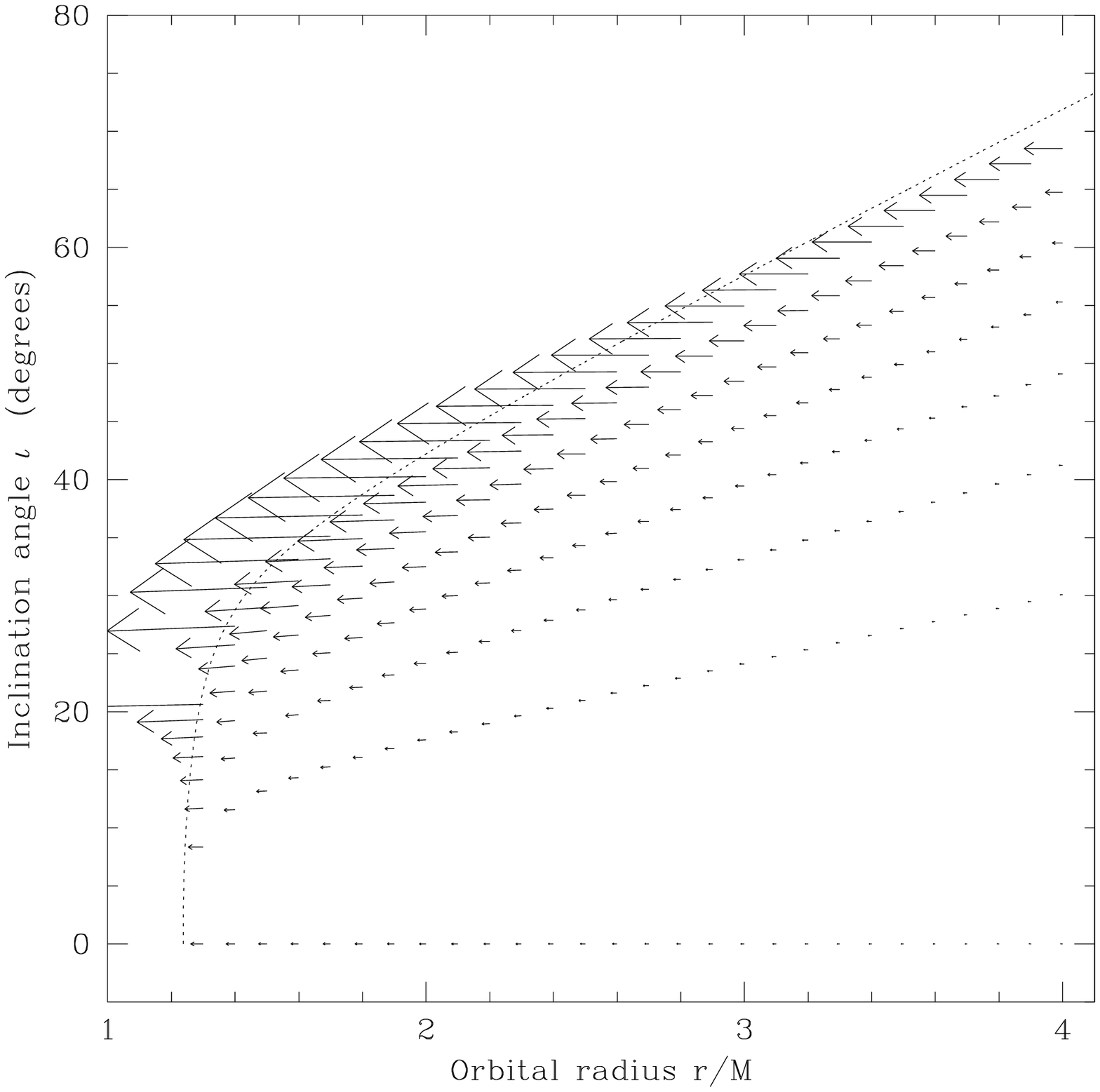, width = 15cm}
\caption{\label{fig:rrfield_0.998}
Circular orbit radiation reaction data near the last stable orbits
(LSO) of a Kerr black hole with $a = 0.998 M$.  Any point $(r,\iota)$
in this plot is a circular geodesic orbit.  The arrows are
proportional to the vector $[(M/\mu)\dot r, (M^2/\mu)\dot\iota]$: the
orientation gives the direction in which gravitational-wave emission
drives the orbit, and the magnitude is proportional to the rate at
which it is so driven.  The dotted line is the LSO --- orbits above
and to the left of this line are dynamically unstable and rapidly
plunge into the black hole.  The arrows get longer as this line is
approached and their stability decreases.  (Additional data were
produced representing radiation reaction for orbits on the LSO; these
orbits are so unstable that their radiation reaction vectors do not
fit on this plot.  These data are used in all computations, however.)
}
\end{center}
\vskip -0.5cm
\end{figure}

An example grid, for a hole with $a = 0.998 M$, is shown in Fig.\
{\ref{fig:rrfield_0.998}}.  The arrows in this plot represent the
vector $(\dot r, \dot\iota)$.  Notice that the evolution gets more
rapid as the LSO (represented by the dotted line) is approached ---
the arrows get longer, until finally the orbit becomes dynamically
unstable and plunges into the hole.  (The diverging vectors on the LSO
have been suppressed in the figure, as they tend to overwhelm the rest
of the data.)  This grid prototypes all the radiation reaction data
used in this paper --- although differing in detail depending on the
black hole spin, they have the general shape and layout shown in Fig.\
{\ref{fig:rrfield_0.998}}.

\subsection{Integrating a trajectory across the grid}
\label{subsec:integrate}

Radiation reaction data on the grid are used to build inspiral
trajectories --- the paths $[r(t),\iota(t)]$ that bodies follow as
they spiral into the black hole.  This is done with Euler's method:
given a starting point $[r_0,\iota_0]$ and a time step $\delta t$,
advance to $[r_0 +\dot r\,\delta t, \iota_0 + \dot\iota\,\delta t]$.
It would be straightforward to use a more sophisticated stepping
algorithm, such as a Runge-Kutta method, to improve the accuracy of
the inspiral trajectory.  For the purpose of a first exploration of
extreme mass ratio inspiral properties, Euler's method is a good
compromise between accuracy on the one hand, and code complexity and
computational cost on the other.  As the trajectory is built, other
useful data are also calculated, particularly the accumulated number
of orbits and the gravitational waveform.

The Euler method integration requires derivatives at arbitrary points
in the orbital parameter space, which in turn requires smooth methods
for interpolating off of the grid points.  Linear interpolations turn
out not to be adequate: discontinuities in the data's second
derivative as grid boundaries are crossed noticeably affects the
phasing of the computed gravitational waveform\footnote{The
discontinuities are amazingly clear when one transforms a
gravitational waveform so produced into audio data.}.  Two dimensional
cubic spline interpolation {\cite{numrec}} seems to produce an
inspiral trajectory that is adequately smooth; it is used for all
interpolations in this analysis.

The spline interpolation is done in the index coordinates, $(i,j)$.
Suppose we wish to interpolate data for some field $\chi$ onto
$(r,\iota)$.  First, we interpolate in $j$ [which maps simply onto
$r$; cf.\ Eq.\ (\ref{eq:discret_rad})] for each value of $i$, yielding
a one-index set of data at the radius $r$:
\begin{equation}
\mbox{Spline in $r$}:\qquad \chi_{ij} \mapsto \chi_i(r)\;.
\label{eq:spline_in_r}
\end{equation}
Since all of the relevant data now live at a single radius, the index
$i$ maps onto $\cos\iota$.  Interpolate in $\cos\iota$ to get the
final data:
\begin{equation}
\mbox{Spline in $\cos\iota$}:\qquad \chi_i(r) \mapsto \chi(r,\iota)\;.
\label{eq:spline_in_cosiota}
\end{equation}
This procedure behaves very well with the radiation reaction data and
waveform data discussed in Sec.\ {\ref{subsec:grid}}.

The procedure for generating the inspiral trajectory and gravitational
waveform now reduces to a simple recipe:
\begin{enumerate}

\item Pick a starting coordinate, $(r_0,\iota_0; t = 0)$.

\item Interpolate the data $d\cos\iota/dt$, $\left[\cos\iota
-\cos\iota_{\rm LSO}\right]\dot r \equiv \dot\rho$, $\Omega_\phi$,
$\Omega_\theta$, ${\cal A}_{lmk}$, and $\Phi_{lmk}$ onto this
coordinate.  Note that for the waveform amplitude and phase, this
requires interpolations for every value of $l$, $m$, and $k$, which
becomes computationally intensive.  As discussed in Sec.\
{\ref{subsec:grid}}, $\Phi_{lmk}$ is not stored on the equator when
$k\ne 0$.  One gets data for $\Phi_{l,m,k\ne0}$ between the equator
and the next grid point by extrapolation.  This works very well
because, at constant radius, $\Phi_{lmk}$ is nearly flat as a function
of $\cos\iota$.

\item Generate the gravitational waveform at that moment on the
trajectory:
\begin{eqnarray}
\omega_{mk} &=& m\Omega_\phi + k\Omega_\theta\;,
\nonumber\\
h(t) &\equiv& h_+(t) - i h_\times(t) =
\sum_{l = 2}^{l_{\rm max}}\sum_{m = -l}^l
\sum_{k = -k_{\rm max}}^{k = k_{\rm max}}
{{\cal A}_{lmk}\over\omega_{mk}^2}{_{-2}S}^{a\omega_{mk}}_{lm}(\vartheta)
e^{i\Phi_{lmk}}e^{i m\varphi}e^{-2\pi i(mN_\phi + kN_\theta)}\;.
\label{eq:genwaveform}
\end{eqnarray}
Here, $\vartheta$ is the angle between the observer's line of sight
and the hole's spin axis, and $\varphi$ is the orbit's $\phi$
coordinate at $t = 0$.  Formally, $l_{\rm max} = k_{\rm max} =
\infty$.  In practice, these numbers are truncated at something much
smaller.  The last term in $h(t)$ is a harmonic of the accumulated
orbital phases; the factor in the exponential is equivalent to $-i\int
dt(m\Omega_\phi + k\Omega_\theta)$.  It would reduce to $-2\pi i
(m\Omega_\phi + k\Omega_\theta)t$ if the orbital frequencies did not
themselves evolve.

\item Compute the number of orbits accumulated so far:
\begin{eqnarray}
N_\phi(t + \delta t) &=& N_\phi(t) + \Omega_\phi\,\delta t/2\pi\;,
\nonumber\\
N_\theta(t + \delta t) &=& N_\theta(t) + \Omega_\theta\,\delta t/2\pi\;.
\label{eq:num_orbits}
\end{eqnarray}
Since the frequencies of $\phi$ and $\theta$ motion are generally
incommensurate, these two numbers can be quite different.

\item Take an Euler step to a new coordinate:
\begin{eqnarray}
t_{\rm new} &=& t_{\rm old} + \delta t\;,
\nonumber\\
r_{\rm new} &=& r_{\rm old} + {1\over\cos\iota
- \cos\iota_{\rm LSO}}\dot\rho\,\delta t\;,
\nonumber\\
\cos\iota_{\rm new} &=& \cos\iota_{\rm old} + {d\cos\iota\over dt}\,
\delta t\;.
\label{eq:takeastep}
\end{eqnarray}

\item Go to Step 2 and repeat.  Continue until the inspiraling body
crosses the LSO.

\end{enumerate}

Results from applying this analysis to a large number of inspirals of
black holes of various spins are discussed next.

\section{Results: Inspiral trajectories}
\label{sec:trajectories}

Before generating inspiral trajectories and gravitational waveforms,
we must choose a sample of black hole spins --- the trajectories are
unique for each spin choice of the massive black hole.  (The effect of
varying other parameters, particularly the mass and mass ratio, can be
obtained by scaling.)  We would like a range of $a$ that covers at
least qualitatively the range likely in astrophysical extreme mass
ratio inspirals.

This analysis was done for four representative black hole spins: $a =
0.998 M$, $a = 0.3594M$, $a = 0.8M$, and $a = 0.05M$.  The first two
were selected because they are rigorously calculable.  The value $a =
0.998M$ is the ``astrophysically maximal'' spin one finds when the
black hole evolves due to thin disk accretion: preferential capture of
counter-rotating photons versus co-rotating photons buffers the spin
at $0.998M$, preventing the hole from reaching the Kerr maximal value
$a = M$ {\cite{kip74}}.  The second choice is obtained by locking the
rotation frequency of the horizon, $\Omega_H = a/2Mr_+$, to the
orbital frequency of the innermost equatorial orbit at $r_{\rm pro}$,
$\Omega_\phi = M/[r_{\rm pro}^{3/2} + aM^{1/2}]$.  Equating these two
frequencies and using Eq.\ (\ref{eq:r_pro}) yields $a = 0.3594M$.
This spin might be obtained if at some point in the black hole's
history strong magnetic fields threaded the event horizon and the
inner edges of an accretion disk, torquing the hole such that it
became locked to the disk's rotation {\cite{roger_comment}}.  The
third choice, $a = 0.8M$, is primarily chosen because it breaks up the
large gap between $0.998M$ and $0.3594M$.  It is worth noting that
this spin is in the range predicted by detailed models of black hole
evolution in the presence of magnetohydrodynamic torque, such as are
described in some models of quasar engines {\cite{moderskietal}}.
Finally, $a = 0.05M$ is chosen to give an example of inspiral into a
slowly spinning black hole.  In all cases, the numbers presented are
for a $\mu = 1\,M_\odot$ body spiraling into an $M = 10^6\,M_\odot$
black hole.

Results are summarized in Figs.\ {\ref{fig:traj_H_0.998}},
{\ref{fig:traj_nH_0.998}}, {\ref{fig:traj_0.3594}},
{\ref{fig:traj_0.8}}, and {\ref{fig:traj_0.05}}.  Before discussing
each case, it is worth noting some general trends in the data.  The
span of data in all cases has been chosen so that the total inspiral
lasts about $630 - 650$ days for the shallowest inclination angles.
Inspiral typically halts when the small body crosses the LSO; at the
very shallowest inclination angles, it halts when the span of
radiation reaction data ends (very close to the LSO).

The radius of the LSO as a function of inclination angle, $r_{\rm
LSO}(\iota)$, depends quite strongly on the black hole spin ---
consider that for $a = 0$, $r_{\rm LSO}(\iota) = 6M$ independent of
$\iota$, whereas for $a = M$, $r_{\rm LSO}(0^\circ) = M$, $r_{\rm
LSO}(180^\circ) = 9M$.  Because of this, the starting radius of the
inspiral is moved out to larger values as $a$ is decreased.  Also, the
total number of starting inclination angles is less for large $a$:
when $a$ is close to $M$, trajectories that begin at high values of
$\iota$ hit the LSO more quickly.  Thus, in the data discussed below,
inspiral into small $a$ black holes occurs at larger coordinate radius
than for large $a$, and more starting inclination angles are included
in those data sets.

In all cases, I show the accumulated inspiral time, the number of
orbits about the spin axis ($N_\phi$) and the accumulated number of
oscillations in $\theta$ ($N_\theta$).  These numbers are easily
scaled to different masses and mass ratios: the trajectory shapes are
independent of the masses (provided the radial axis is $r/M$), and
\begin{equation}
T_{\rm inspiral} \propto {M^2\over\mu}\;,\quad
N_{\phi,\theta} \propto {M\over\mu}\;.
\label{eq:time_and_N_scaling}
\end{equation}
The trajectories shown here thus apply for any black hole mass and any
extreme mass ratio.

\subsection{Trajectories for $a = 0.998M$}
\label{subsec:traj_0.998}

\begin{figure}[ht]
\begin{center}
\epsfig{file = 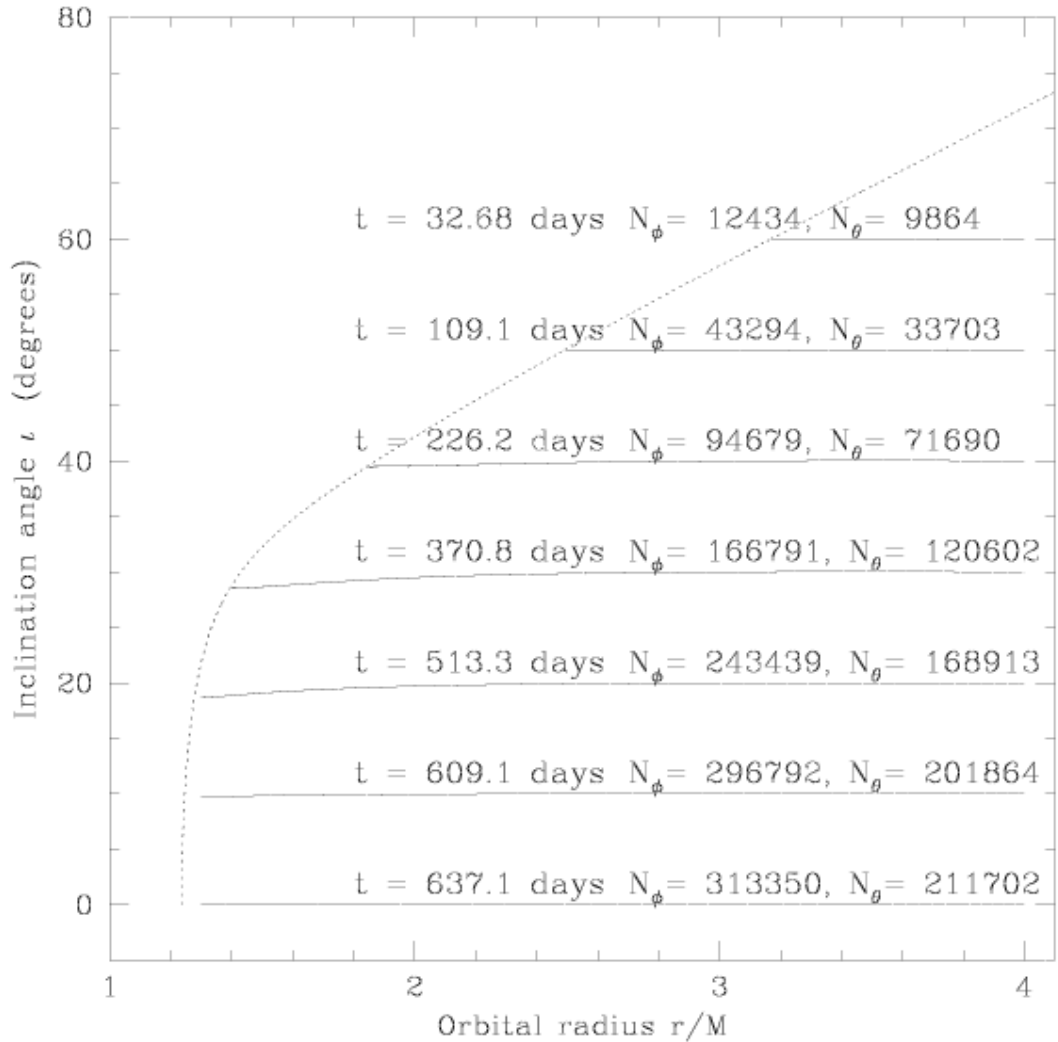, width = 15cm}
\caption{\label{fig:traj_H_0.998}
Inspiral trajectories in the strong field of a Kerr black hole with $a
= 0.998M$.  To make this plot, the data shown in Fig.\
{\ref{fig:rrfield_0.998}} were integrated using the procedures
discussed in Sec.\ {\ref{subsec:integrate}}, assuming that the black
hole has mass $M = 10^6\,M_\odot$ and that the inspiraling body has
mass $\mu = 1\,M_\odot$.  The trajectory shapes are independent of the
two masses, so the inspiral times and accumulated number of cycles
$N_{\phi,\theta}$ can be rescaled to other masses quite easily:
$T_{\rm inspiral} \propto M^2/\mu$, $N_{\phi,\theta} \propto M/\mu$.
Notice that the trajectories are nearly flat --- $\iota$ decreases,
but not very much, over these inspirals.
}
\end{center}
\vskip -0.5cm
\end{figure}

\begin{figure}[ht]
\begin{center}
\epsfig{file = 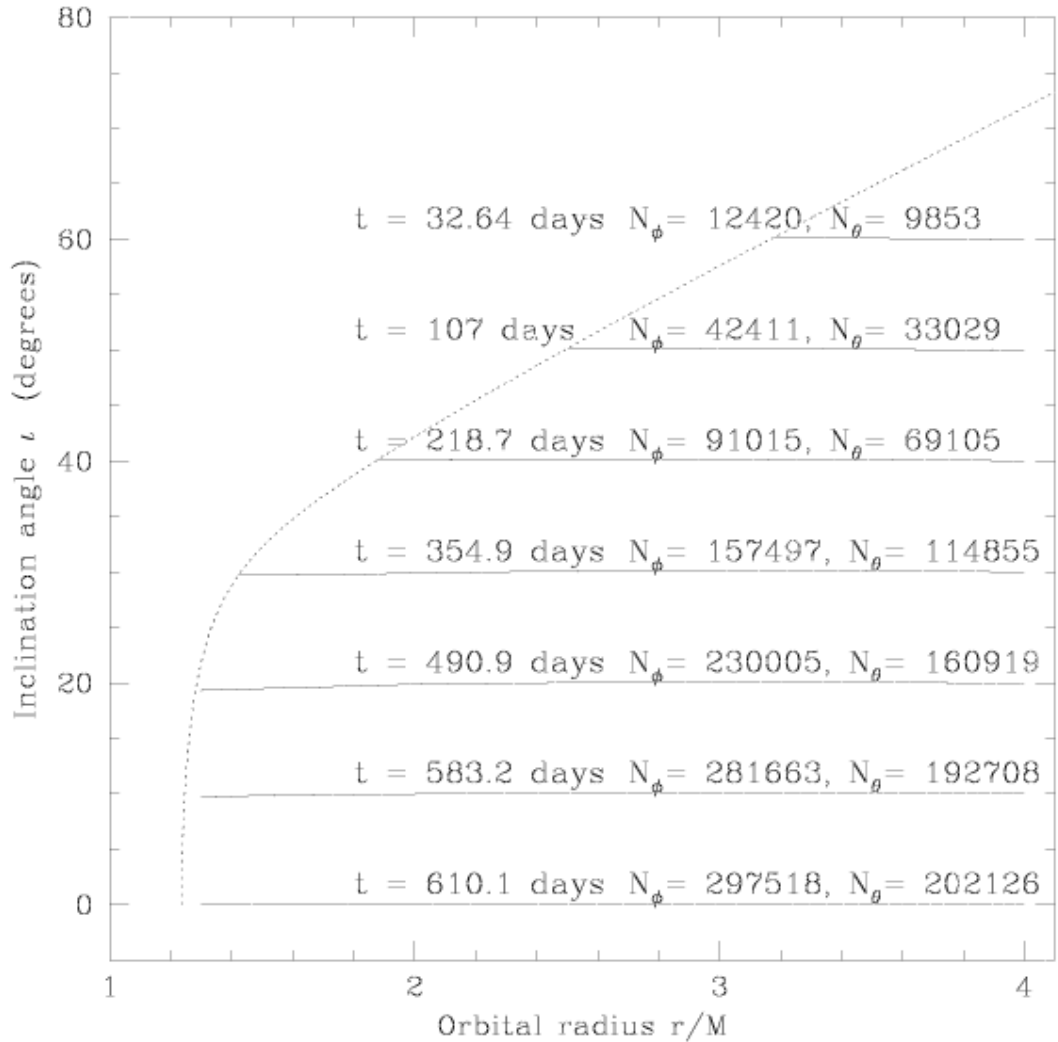, width = 15cm}
\caption{\label{fig:traj_nH_0.998}
Inspiral trajectories in the strong field of a Kerr black hole with $a
= 0.998M$, ignoring the flux down the horizon --- cf.\ Eq.\
(\ref{eq:turnHflux_onoff}), with $\eta = 0$.  The figure is otherwise
identical to Fig.\ {\ref{fig:traj_H_0.998}}.  The trajectory shapes
change very slightly ($\iota$ does not decrease quite as much over the
inspiral), though that effect is very small.  Much more interestingly,
the inspiral is markedly {\it faster}: especially at shallow
inclination angle, the small body takes several weeks less spiraling
to the LSO, and executes many thousands fewer orbits.  As discussed in
Sec.\ {\ref{subsec:traj_0.998}}, this illustrates how tidal coupling
between the small body and the event horizon strongly impacts the
inspiral: a tidal bulged is raised on the hole, which, due to the
hole's rapid rotation in this case, transfers rotational kinetic
energy to the small body's orbit.
}
\end{center}
\vskip -0.5cm
\end{figure}

\begin{figure}[ht]
\begin{center}
\epsfig{file = 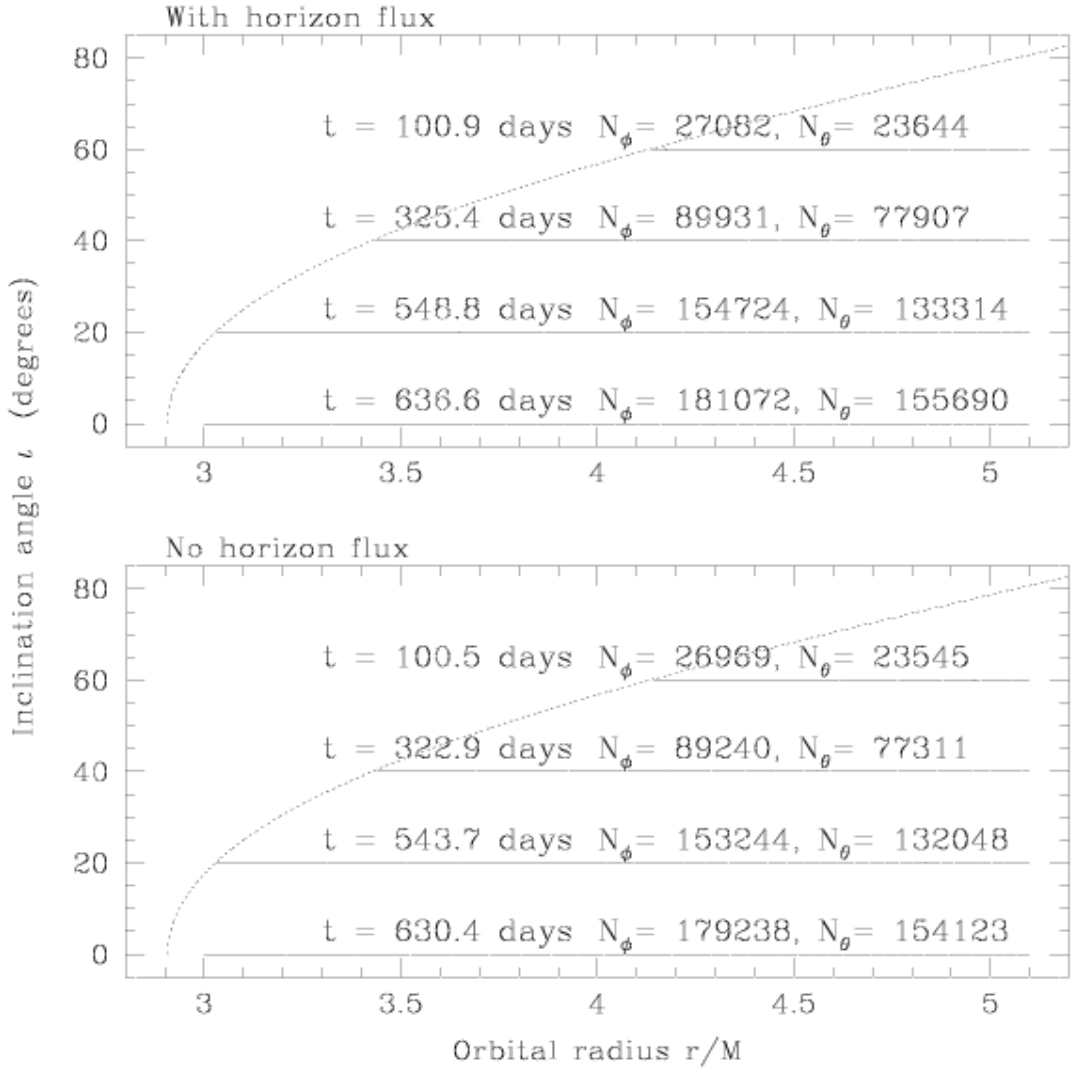, width = 15cm}
\caption{\label{fig:traj_0.8}
Inspiral trajectories in the strong field of a Kerr black hole with $a
= 0.8M$.  The top panel includes the effects of down-horizon flux; the
bottom panel does not.  The span of data is chosen so that the the
total inspiral duration is similar to that shown in Fig.\
{\ref{fig:traj_H_0.998}}.  As when $a = 0.998M$, inspiral is quicker
when the horizon flux is ignored.  However, the magnitude of the
effect is much smaller.  The torque on the orbit that the horizon's
tidal bulge exerts is not as great because the hole rotates less
quickly.  Also, orbits do not come as close to the horizon for $a =
0.8M$ as they do for $a = 0.998M$.
}
\end{center}
\vskip -0.5cm
\end{figure}

Trajectories for inspiral into a hole with spin $a = 0.998M$ are
summarized in Fig.\ {\ref{fig:traj_H_0.998}}.  This figure shows
inspiral from $r = 4M$ into the LSO for several starting values of the
inclination angle $\iota$.  The accumulated inspiral time, orbits about
the spin axis, and number of oscillations in $\theta$ label each
trajectory.

Notice that the trajectories are nearly flat --- $\iota$ barely
changes, decreasing somewhat as the small body passes through the very
strong field.  This decrease is contrary to weak-field expectations
{\cite{paperI,ryan_circ}}, and arises because the characteristics of
Kerr geodesics become locked to the event horizon in the very strong
field; see Ref.\ {\cite{nhs}} for detailed discussion.  The fact that
the change in $\iota$ is so small is very interesting.  The change
turns out to be even smaller when $a$ is not so large.

Figure {\ref{fig:traj_H_0.998}} illustrates the true strong-field
inspiral sequences predicted by general relativity, including the
effects of radiation out to infinity and down the event horizon.  An
interesting experiment is to ``turn off'' the radiation down the
horizon, setting the parameter $\eta$ in Eq.\
(\ref{eq:turnHflux_onoff}) to 0.  The inspiral trajectories found in
this exercise are plotted in Fig.\ {\ref{fig:traj_nH_0.998}}.  Aside
from turning off the horizon flux, these trajectories are computed
with conditions identical to the trajectories plotted in Fig.\
{\ref{fig:traj_H_0.998}}.  Notice that, particularly for shallow
inclination angle ($\iota \alt 50^\circ$), the small body inspirals
{\it more quickly} when the flux down the horizon is not included.
The effect is quite significant: inspiral of the shallowest orbits is
shortened by several weeks, orbiting $\sim10^4$ fewer times when the
horizon flux is ignored.

That the horizon flux tends to prolong the inspiral is, on first
consideration, extremely surprising.  Intuitively, one would guess
that each flux would be an energy sink, so that each reduces the
orbit's energy.  Because of the hole's rapid rotation, this is not the
case: the phenomenon of superradiance plays a major role.
Superradiance is essentially a manifestation of the Penrose process
--- a radiation pulse incident on the black hole is backscattered with
a gain in the pulse's energy {\cite{pressteuk}}.  The energy gain is
at the expense of the hole's rotation --- rotational kinetic energy is
converted to radiative energy.  Thus, when the hole spins rapidly
enough, the horizon is a {\it source} of energy, not a sink.

A rather more physical picture of this phenomenon (which explains why
the energy is transferred to the orbit, not simply radiated to
infinity) can be put together based on work by James Hartle
{\cite{hartle}}.  The portion of the Weyl tensor which describes
radiation down the event horizon can be understood as the tidal field
of the orbiting body acting on the black hole.  This field distorts
the horizon, raising a tidal bulge on the hole.  The distortion can be
quantified by computing the curvature of the horizon.  Using formulas
from Ref.\ {\cite{hartle}}, it is easy to show that the distortion is
given by a sum over the coefficients $Z^\infty_{lmk}$ that in this
analysis describe the radiation flux at the horizon.

The black hole's tidal bulge is equivalent to the bulge that the moon
raises on the earth.  When the rotation frequency of the hole is
greater than the $\phi$ frequency, $\Omega_H > \Omega_\phi$, the
hole's spin drags the bulge {\it ahead} of the orbiting body, in
exactly the manner in which a tidal bulge on a viscous, fluid body is
dragged by the body's rotation.  (One can in fact interpret the
horizon's behavior in terms of a viscous membrane {\cite{membrane}}.)
In a reference frame that co-rotates with the small body's orbit, the
bulge appears to lead the small body by some angle.  The bulge thus
exerts a torque that acts to increase the body's orbital velocity,
transferring the hole's rotational energy into the orbit and slowing
the inspiral.  A very accessible discussion of this point (which
directly relates the ``tidal bulge'' effect to a torque) is given in
Chapter VII of Ref.\ {\cite{membrane}}; the fundamental underpinnings
of this analysis are developed in Refs.\
{\cite{hawkinghartle,hartle}}.

\begin{figure}[ht]
\begin{center}
\epsfig{file = 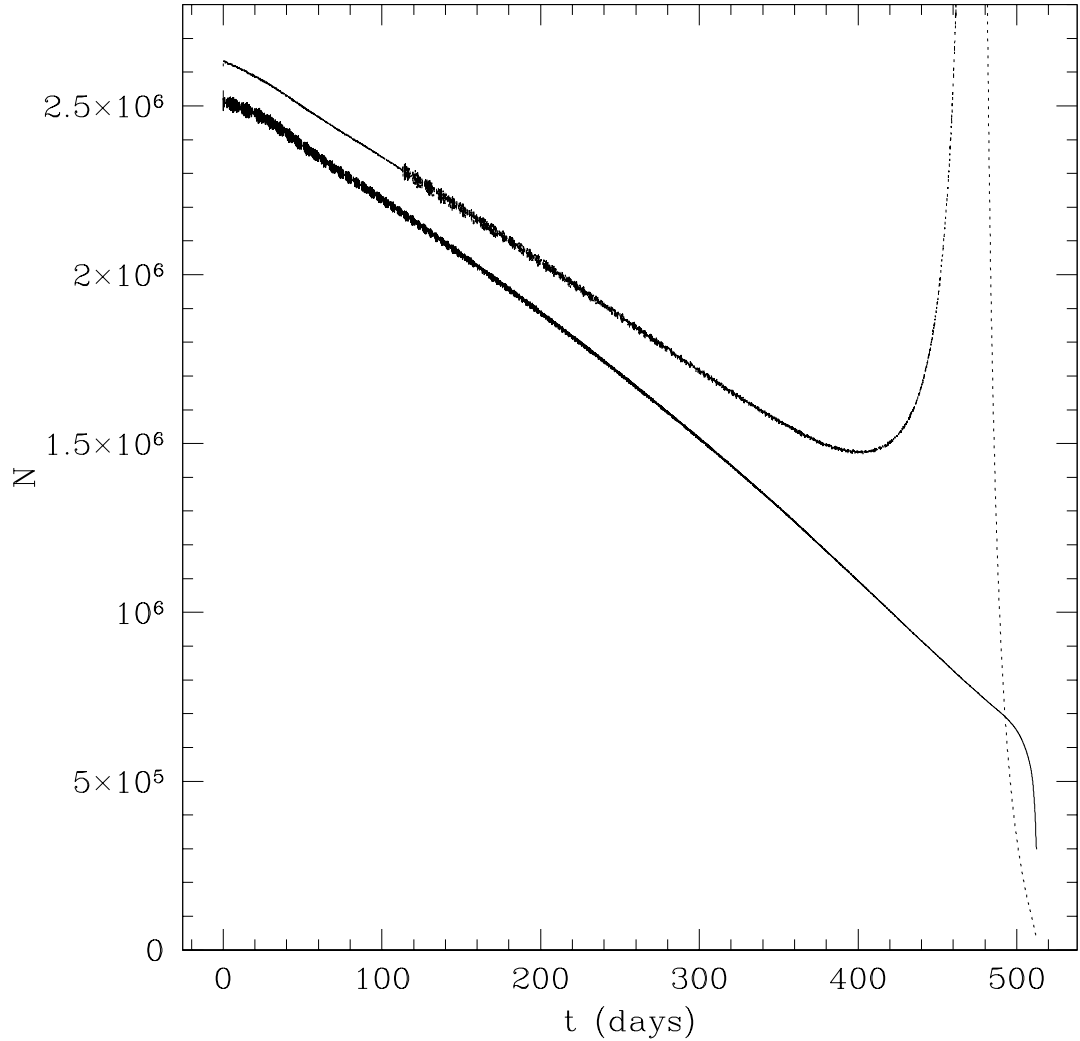, width = 15cm}
\vskip 0.5cm
\caption{\label{fig:adiabat}
The evolution of the adiabaticity parameters ${\cal N}_{\phi,\theta} =
\Omega_{\phi,\theta}^2/2\pi{\dot\Omega}_{\phi,\theta}$ for the inspiral
track beginning at $\iota = 20^\circ$ shown in Fig.\
{\ref{fig:traj_H_0.998}}.  The solid line is ${\cal N}_\phi$, the
dotted line ${\cal N}_\theta$.  Except at the extreme end, ${\cal
N}_{\phi,\theta}\gg 1$, indicating that the inspiral is indeed
adiabatic, as required.  Note that ${\cal N}_{\phi,\theta}\propto
M/\mu$, indicating that the adiabatic requirements are unlikely to be
met if $M/\mu < 1000$.  The divergence in ${\cal N}_\theta$ is due
to ${\dot\Omega}_\theta$ passing through zero and changing sign during
the inspiral --- it ceases chirping up, and begins chirping down.
}
\end{center}
\vskip -0.5cm
\end{figure}

As discussed in Sec.\ {\ref{subsec:circ_radreact}}, it is important to
monitor the adiabaticity parameters ${\cal N}_\phi$ and ${\cal
N}_\theta$ during the inspiral.  The values of ${\cal N}_\phi$ and
${\cal N}_\theta$ for the inspiral beginning at $\iota = 20^\circ$ are
plotted in Fig.\ {\ref{fig:adiabat}}; ${\cal N}_\phi$ is plotted as
the solid line, ${\cal N}_\theta$ is the dotted line.  We see the
adiabaticity condition ${\cal N}_{\phi,\theta} \gg 1$ is strongly
satisfied everywhere except right before the end of inspiral.  (This
is because the inspiral rate sweeps up as inspiral proceeds, and can
be regarded as a precursor to the final plunge.)  The adiabaticity
parameters scale with $M/\mu$.  This is a warning that the results of
this code cannot be taken seriously if the mass ratio is not extreme
enough: when $M/\mu\alt 1000$, the final stages of inspiral will not
be adiabatic.  An interesting feature in this plot is the divergence
in ${\cal N}_\theta$ near $t = 470\,{\rm days}$.  This divergence
occurs because ${\dot\Omega}_\theta$ switches sign there: the
frequency stops increasing and begins decreasing.  This will be
discussed at greater length in Sec.\ {\ref{sec:waveforms}}.

The adiabaticity parameters appear qualitatively as in Fig.\
{\ref{fig:adiabat}} in all cases studied here (with the modification
that the divergence in ${\cal N}_\theta$ only occurs for $a =
0.998M$), so no other examples of ${\cal N}_{\phi,\theta}$ will be
shown.

\subsection{Trajectories for $a = 0.3594M$}
\label{subsec:traj_0.3594}

\begin{figure}[ht]
\begin{center}
\epsfig{file = 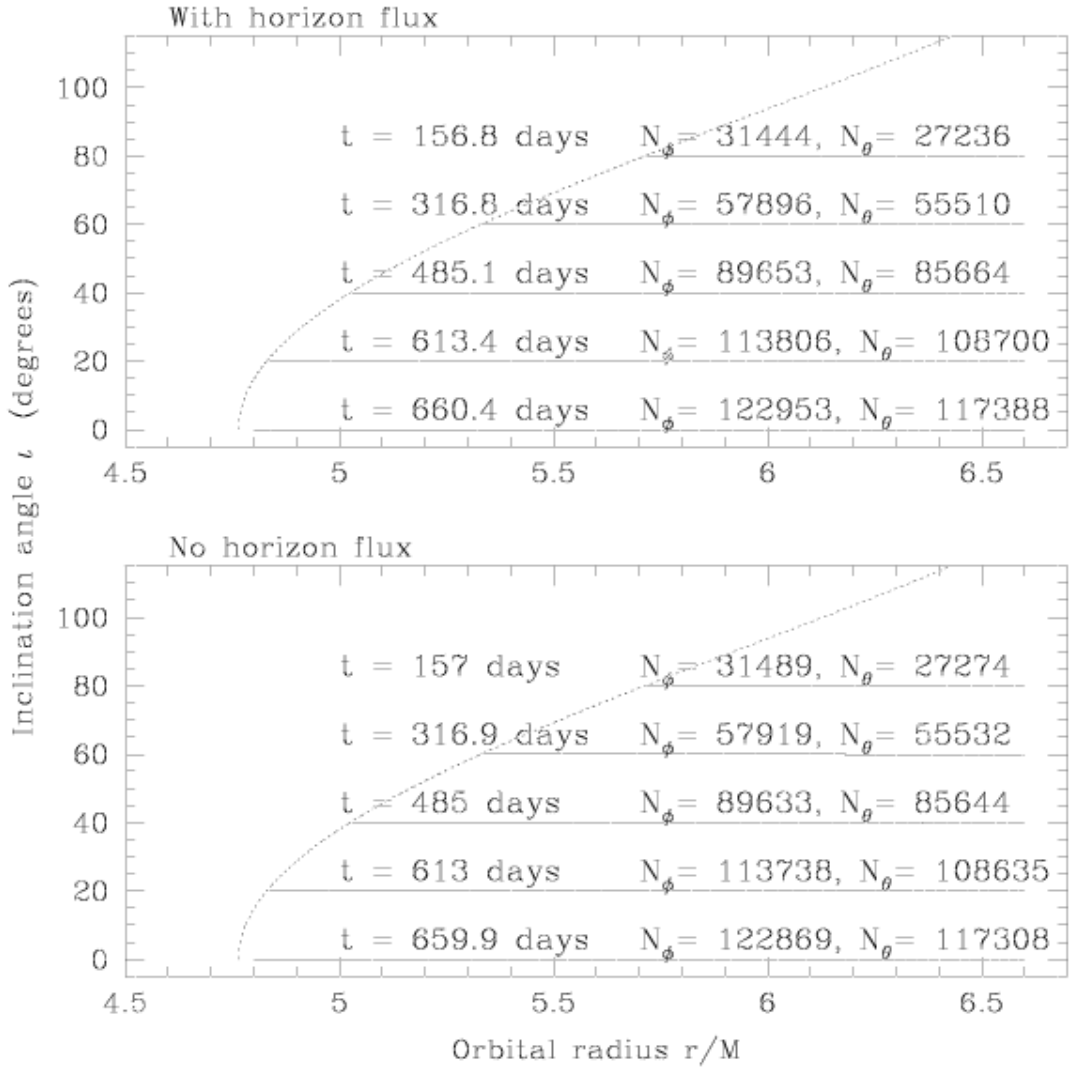, width = 15cm}
\caption{\label{fig:traj_0.3594}
Inspiral trajectories in the strong field of a Kerr black hole with $a
= 0.3594M$.  The top panel includes the effects of the down-horizon
flux; the bottom panel does not.  The span of data is chosen so that
the total inspiral duration is similar to that shown in Fig.\
{\ref{fig:traj_H_0.998}}.  For this spin, inspiral is nearly identical
with the horizon flux included or disincluded: inspiral is slightly
faster without horizon flux, but not nearly so much faster as when $a
= 0.998M$.  This is largely because the spin is not fast enough to
drag the tidal bulge on the horizon so far forward.  In fact, at the
innermost orbits the bulge should be essentially in perfect face-on
lock with the orbiting body, since $a = 0.3594M$ is the spin value at
which the horizon spin frequency matches the innermost orbital
frequency.
}
\end{center}
\vskip -0.5cm
\end{figure}

Trajectories for inspiral into a black hole with spin $a = 0.3594M$
are summarized in Fig.\ {\ref{fig:traj_0.3594}}.  These trajectories
are set up in a similar manner as those for $a = 0.998M$; notable
differences are that they begin at $r = 6.6 M$, and more starting
values of $\iota$ are included.  The trajectories are even more flat
in this case than they are when $a = 0.998M$.  The inclination angle
actually increases slightly in all cases (as weak-field analyses
predict), but the amount of increase is too small to be noticeable on
the figure.

Turning off the horizon flux does not have a very marked effect on
inspiral for this spin, as can be seen by comparing the upper panel
(horizon flux included) and lower panel (horizon flux ignored) of
Fig.\ {\ref{fig:traj_0.3594}}. This is essentially by construction:
the spin $a = 0.3594M$ is the value for which the horizon spin
frequency matches the innermost orbital frequency.  The lead angle
between the tidal bulge and the orbiting body ranges from very small
to nonexistent --- as the body spirals through the innermost orbits,
the bulge is nearly perfectly lined up with the body, exerting almost
no torque on it.  The small difference between the trajectories in the
two panels is not surprising.

\subsection{Trajectories for $a = 0.8M$}
\label{subsec:traj_0.8}

Inspiral trajectories for a black hole with $a = 0.8M$ are shown in
Fig.\ {\ref{fig:traj_0.8}}.  Not surprisingly, the properties of these
trajectories are intermediate to those shown for the cases $a =
0.998M$ and $a = 0.3594M$.  As in the case $a = 0.3594M$, the change
in inclination angle is extremely small, increasing slightly.

Inspiral is faster when the horizon flux is not included.  This,
again, is not a surprise: we expect that when $a = 0.8M$ the bulge
raised on the event horizon leads the orbiting body, so that tidal
coupling transfers energy from the hole's spin to the orbit.  Although
qualitatively the effect is the same as for the case $a = 0.998M$,
quantitatively the effect is much smaller.  The fractional change in
the inspiral time is no more than about $1\%$ when $a = 0.8M$, as
opposed to $4.2\%$ when $a = 0.998M$.  Part of this is simply because
the spin frequency of the hole is smaller when $a = 0.8M$ --- the
tidal bulge does not lead the orbiting body quite as much.  Also the
body's orbits never come as close to the event horizon as they do when
$a = 0.998M$ --- they reach the dynamical instability before they get
so close.  As a consequence, the tidal coupling is never as strong,
and the integrated effect on the trajectory is relatively small.

\subsection{Trajectories for $a = 0.05M$}
\label{subsec:traj_0.05}

Figure {\ref{fig:traj_0.05}} shows trajectories for inspiral into a
hole with $a = 0.05M$.  The dependence of the inspiral properties on
inclination angle is weak compared to the other cases examined, though
definitely present.  The change in the inclination angle is
practically unnoticeable; the largest change (for inspiral near
$90^\circ$) is $\delta\iota \simeq 0.007^\circ$.

In this case, inspiral is quicker when the horizon flux is included.
The horizon functions as a sink of energy, in accord with simple
intuition: when the horizon flux is not included, the small body does
not spiral in so quickly.  Because the hole rotates so slowly, the
tidal bulge raised on the horizon lags the orbiting body, and the
torque that it exerts on the orbit tends to increase its inspiral
rate.  The magnitude of the effect is quite small (fractional change
in inspiral time is about $0.1\%$).

\section{Results: Gravitational waveforms}
\label{sec:waveforms}

\begin{figure}[ht]
\begin{center}
\epsfig{file = 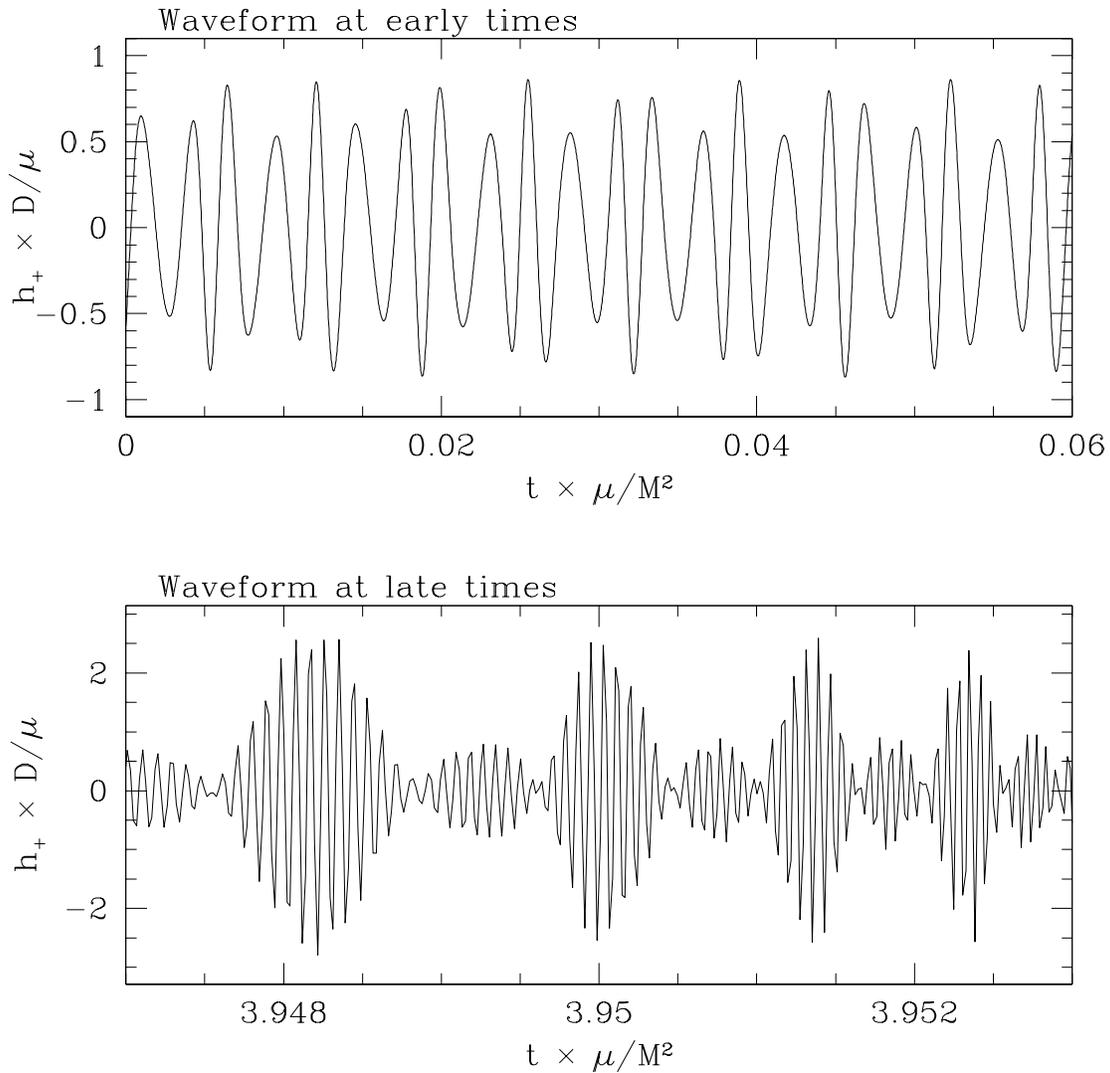, width = 15cm}
\caption{\label{fig:waveform}
The $+$ polarization of the gravitational waveform for the inspiral
trajectory that begins at $\iota = 40^\circ$ about the $a = 0.998M$
black hole, viewed in the hole's equatorial plane.  The upper panel is
the waveform at very early times; the lower panel shows the waveform
shortly before the inspiraling body plunges into the hole.  Notice the
very different time scales in the upper and lower panels.  This is
because of the ``chirping'' evolution of the frequencies $\Omega_\phi$
and $\Omega_\theta$ --- at late times they are quite a bit larger than
they are early on.  Many more orbits per unit time are executed late
in the inspiral than early.
}
\end{center}
\end{figure}

As the inspiral trajectory is generated, the gravitational waveform is
constructed using Eq.\ (\ref{eq:genwaveform}).  An example waveform is
shown in Fig.\ {\ref{fig:waveform}}.  This is the $+$ polarization
generated during the inspiral beginning at $\iota = 40^\circ$ into a
hole with $a = 0.998M$; cf.\ Fig.\ {\ref{fig:traj_H_0.998}}.  It is
viewed in the hole's equatorial plane ($\theta = \pi/2$).  The mass
ratio used here is $10^{-4}$; recall that $10^{-6}$ was used to make
Fig.\ {\ref{fig:traj_H_0.998}}.  The increase was to speed up the
inspiral so that the amount of data generated in the wave was kept
manageable.  The waveform contains contributions from $l = 2, 3, 4$,
and $k = -4$ to $4$.

Some of the interesting features of the waveform are apparent in Fig.\
{\ref{fig:waveform}}.  The modulation of the ``carrier'' signal by
orbital motion in $\theta$ is plain, as is the evolution of the signal
frequencies.  Some of these features are made even clearer by
transforming the gravitational waveform into audio data and playing
the gravitational-wave ``sound''.  I have placed sounds corresponding
to such gravitational-wave signals at the URL listed as Ref.\
{\cite{soundsonweb}}; the reader is invited to play these sounds and
judge for themselves how clear are the various inspiral features.  (In
order for the sounds to lie within the frequency band to which human
ears are sensitive --- and to keep the signal duration reasonable ---
the mass of the black hole is scaled down to $M \sim 100\,M_\odot$.
Discussion of this and other technical points is given on the page
listed in {\cite{soundsonweb}}.)

Extreme mass ratio gravitational waveforms are very usefully described
as {\it multi-voice chirps}.  Strong radiation is emitted at multiple
harmonics of the inspiral frequencies; each harmonic plays the role of
a separate ``voice'' in the ``chorus'' that is the overall waveform.
This structure is quite apparent in the audio versions of the
waveforms.  It can be made clearer by rewriting the waveform
Eq. (\ref{eq:waveform}) as follows:
\begin{equation}
h(t;\vartheta,\varphi) =\sum_{l = 2}^{l_{\rm max}}
\sum_{m = -l}^l\sum_{k = -k_{\rm max}}^{k_{\rm max}}
{\cal H}_{lmk}(t;\vartheta)\exp\left[i\Psi_{lmk}(t;\varphi)\right]\;.
\label{eq:multivoicewave}
\end{equation}
Comparing with Eq.\ (\ref{eq:genwaveform}), we read off the values of
the wave amplitude ${\cal H}_{lmk}$ and phase $\Psi_{lmk}$ in terms of
quantities directly computable from the Teukolsky equation:
\begin{eqnarray}
{\cal H}_{lmk}(t;\vartheta) &=& {{\cal A}_{lmk}
{_{-2}S}^{a\omega_{mk}}_{lm}(\vartheta)/\omega_{mk}^2}\;.
\label{eq:H_def}\\
\Psi_{lmk}(t;\varphi) &=& \Phi_{lmk} + m\varphi
- \int_0^t dt'\left(m\Omega_\phi + k\Omega_\theta\right)\;.
\label{eq:Psi_def}
\end{eqnarray}
[Recall Eq.\ (\ref{eq:amp_phase_def}), defining ${\cal A}_{lmk}$ and
$\Phi_{lmk}$.]  The last line emphasizes that a big portion of the
phase accumulated is nothing more than the integrated orbital phases.

Presenting the waveform in this way gives a very good sense of which
harmonics contribute strongly to a gravitational-wave measurement.  As
will be discussed below, the phase evolution for each voice is very
simple, though their summed effect is typically rather complicated
(cf.\ the summed waveform shown in Fig.\ {\ref{fig:waveform}}).  This
suggests that searching for each individual voice may be more
effective than searching for the entire waveform.  An effective way of
implementing such a search might be developed using the ``Fast Chirp
Transform'' (FCT) of Jenet and Prince {\cite{ricktom}}.  The FCT is a
generalization of the Fast Fourier Transform (FFT; see, e.g., Ref.\
{\cite{numrec}} and references therein).  The FFT provides an
efficient computational implementation of a Fourier transform --- a
decomposition of data onto basis functions $e^{2\pi ift}$.  The FCT
generalizes this to decompose data onto functions whose phases do not
vary linearly with respect to time.  The phase behavior of the basis
functions $e^{i\Phi_{\rm FCT}(t)}$ must be specified by some model;
indications are that it works well for a wide-variety of smoothly
varying phase behaviors.  Although further investigation is needed to
test this idea, the voices of extreme mass ratio inspiral are good
candidates for detection with the FCT.

The results here are restricted to $a = 0.998M$ and $a = 0.3594M$.
The results for these spins contain enough interesting structure to
draw some useful conclusions about extreme mass ratio gravitational
waves.

\subsection{Waveforms for $a = 0.998M$}
\label{subsec:waves_0.998}

Discussion in this section will focus on waves generated during an
inspiral that starts at $\iota = 20^\circ$ (cf.\ Fig.\
{\ref{fig:traj_H_0.998}}).  All data shown corresponds to waves
measured in the hole's equatorial plane.

\begin{figure}[ht]
\begin{center}
\epsfig{file = 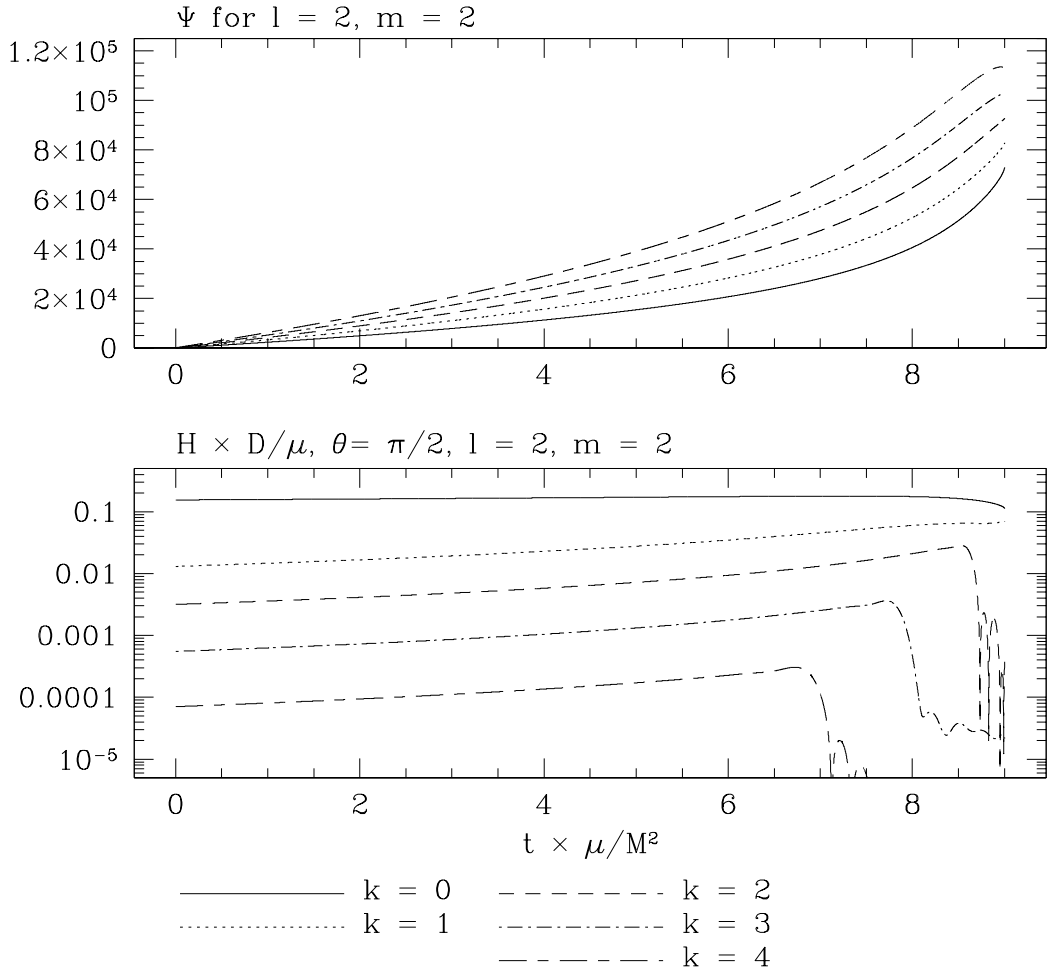, width = 15cm}
\caption{\label{fig:mvc_0.998_l2_m2}
The phase $\Psi_{lmk}$ and amplitude ${\cal H}_{lmk}$ for waveform
harmonics $l = 2$, $m = 2$, and $k$ from 0 to 4, and for spiral into a
hole with $a = 0.998M$.  These amplitudes correspond to measurement in
the hole's equatorial plane.  Note that harmonics other than $k = 0$
become fairly strong at the end of inspiral.  The odd behavior in the
amplitude for $k = 2 - 4$ is because of poor computational resolution:
the crude parameter space resolution used here was not good enough to
accurately capture the change in $Z^H_{lmk}$ at high frequencies as
the body spirals in.
}
\end{center}
\end{figure}

Figure {\ref{fig:mvc_0.998_l2_m2}} shows the amplitudes ${\cal
H}_{lmk}$ and phases $\Psi_{lmk}$ for $l = 2$, $m = 2$, and $k \in [0,
1, 2, 3, 4]$.  Notice that the accumulated phase $\Psi_{22k}$ for all
values of $k$ is quite large --- the number of accumulated radians
ranges from about $8\times10^4$ to $1.1\times10^5$.  The accumulated
phase scales inversely with the mass ratio, so the accumulated phase
in these harmonics would be $\sim 10^7$ radians for a mass ratio of
$10^{-6}$ (as was used in Figs.\ {\ref{fig:traj_H_0.998}},
{\ref{fig:traj_0.3594}}, {\ref{fig:traj_0.8}}, {\ref{fig:traj_0.05}}).
It is by tracking the phase over this large number of radians that it
will be possible to determine the characteristics of the black hole
spacetime with very high accuracy.

The strongest waves are emitted in the $k = 0$ harmonic.  Indeed, the
$l = 2$, $m = 2$, $k = 0$ voice turns out to be the strongest of the
various harmonics that contribute to the gravitational waveform ---
not too surprising, since this is the leading quadrupole component of
the waves.  Note, though, that the $k = 1$ harmonic becomes almost as
strong as the $k = 0$ voice late in the inspiral.  This is a generic
feature of the multi-voice chirps: late in the inspiral, when waves
are emitted from deep in the black hole's strong field, harmonics
which were not initially very important can contribute strongly to the
waves.

The amplitudes corresponding to $k = 2$, $k = 3$ and $k = 4$ ``crash''
and behave rather oddly late in the inspiral.  This is a resolution
problem caused by the crude manner with which I have discretized the
orbital parameter space --- constant Boyer-Lindquist radial steps with
$\delta r = 0.1 M$.  In this region, $p_{mk}\sim (0.2 - 0.5)/M$ and
$|dr^*/dr| \sim 30 - 40$.  Following the discussion near Eq.\
({\ref{eq:suggested_dr}}), the choice $\delta r = 0.1M$ is not really
sufficient to follow the phase variation of $Z^H_{lmk}$; decreasing
the step by a factor of $2$ to $5$ would vastly improve this analysis.

\begin{figure}[ht]
\begin{center}
\epsfig{file = 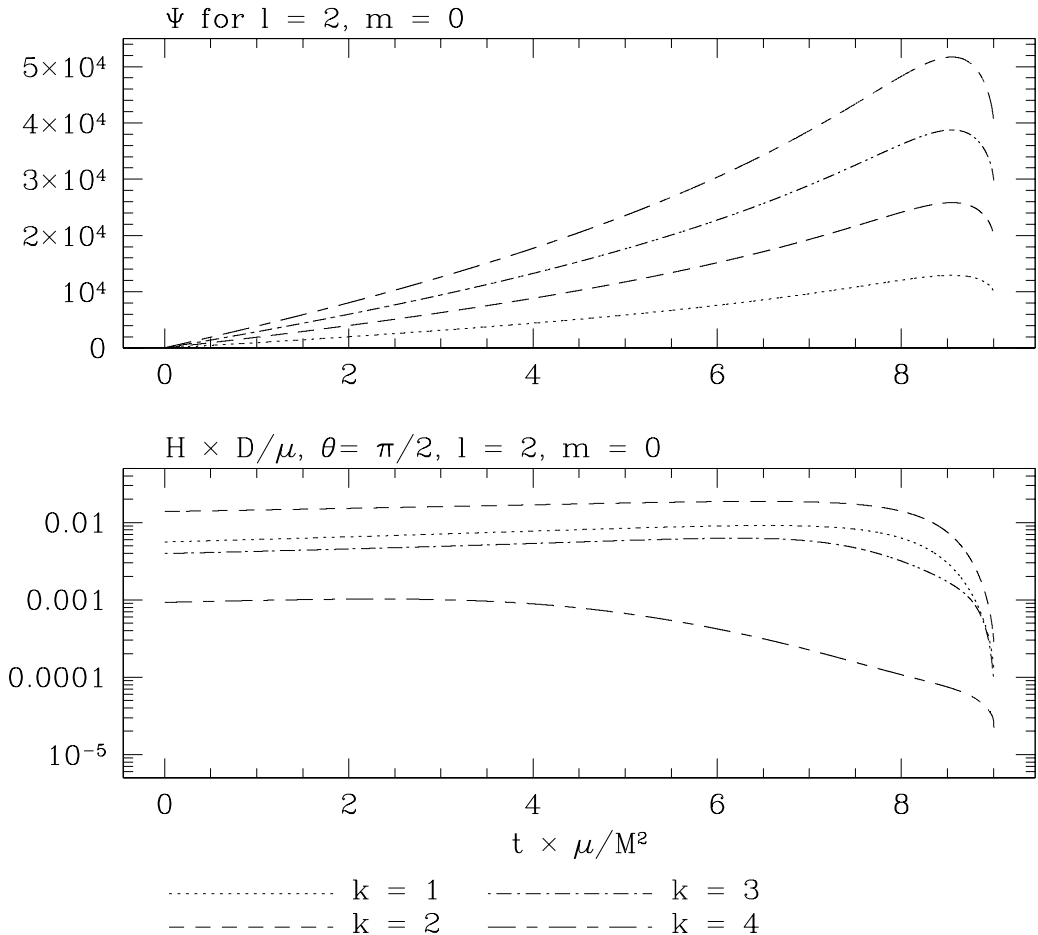, width = 15cm}
\caption{\label{fig:mvc_0.998_l2_m0}
The phase $\Psi_{lmk}$ and amplitude ${\cal H}_{lmk}$ seen in the
equatorial plane, for waveform harmonics $l = 2$, $m = 0$, and $k$
from 0 to 4, and for spiral into a hole with $a = 0.998M$.  In this
case, no frequencies are high enough to cause severe resolution
problems, as was the case for $l = 2$, $m = 2$.  Note that the phase
evolves from high to low late in the inspiral.  This is because the
phase is dominated by the behavior of $\Omega_\theta$, which chirps
backwards when the body gets close to the event horizon; cf.\ Fig.\
{\ref{fig:omegas}}.  This is essentially a gravitational redshifting
effect.  Although this is in principle an interesting signature of the
black hole's strong field, the wave amplitude gets weak as the
frequency gets small.  This is not surprising, since the rate of
change of the binary's multipole moments decreases with the frequency.
}
\end{center}
\end{figure}

\begin{figure}[ht]
\begin{center}
\epsfig{file = 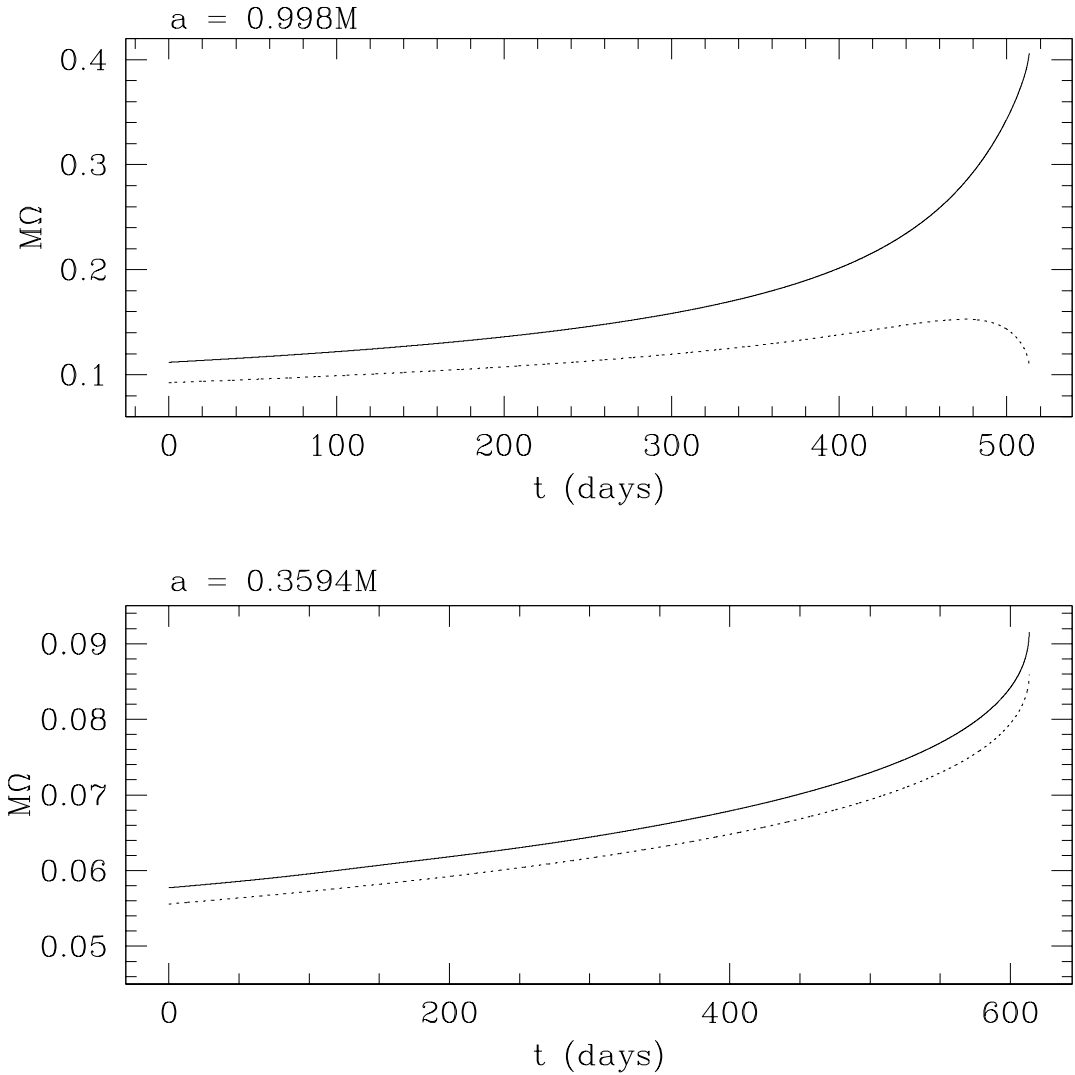, width = 15cm}
\caption{\label{fig:omegas}
Evolution of the orbital frequencies $\Omega_\phi$ (solid line) and
$\Omega_\theta$ (dotted line) for inspiral beginning at $\iota =
20^\circ$.  The upper panel is for inspiral into a black hole with $a
= 0.998M$, the lower for inspiral into a hole with $a = 0.3594 M$.  In
both cases, the behavior of $\Omega_\phi$ is qualitatively ``normal'':
it monotonically chirps upward in a very familiar manner.  In the case
$a = 0.998M$, the behavior of $\Omega_\theta$ is comparatively
unusual.  This is because these orbits are very near to the event
horizon --- their motion in $\theta$ is slowed as seen by distant
observers.  When $a = 0.3594M$, no orbits come close enough to the
horizon for this slowing to have a significant effect.  In fact, all
orbits in the chosen observation band are at relatively large radius
for $a = 0.3594M$, which is why all frequencies are quite smaller than
when $a = 0.998M$ (note the different vertical scales in the two
panels).
}
\end{center}
\end{figure}

The behavior for $l = 2$, $m = 0$ is plotted in Fig.\
{\ref{fig:mvc_0.998_l2_m0}}.  These phases actually {\it decrease} at
late times.  The strongest contributor to the phase evolution of these
waves is the frequency $\Omega_\theta$, which begins decreasing at
late stages of the inspiral.  The evolution of $\Omega_\theta$ and
$\Omega_\phi$ are shown in the upper panel of Fig.\
{\ref{fig:omegas}}.  The decrease in $\Omega_\theta$ occurs because
the small body orbits very close to the event horizon towards the end
of inspiral.  Its $\theta$ motion is highly redshifted as seen by
distant observers.  (The $\phi$ motion, by contrast, is not
redshifted; instead, it limits to the hole's spin frequency as the
orbit becomes locked to the dragging of inertial frames.)  No reverse
chirp occurs when $a = 0.3594M$ --- the inspiraling body is never
close enough to the horizon for there to be a significant redshift
effect.

Measuring a reverse chirp would be in principle a wonderful probe of
the black hole's strong field properties, since it can only occur for
orbits that get very close to the event horizon.  Unfortunately, the
reversal in the phase evolution coincides with a rather rapid decrease
in amplitude for these voices, as can be seen by comparing the upper
and lower panels of Fig.\ {\ref{fig:mvc_0.998_l2_m0}}.  This is
because the relevant source multipole moments vary more slowly as the
frequency drops.  Since this variation sets the wave amplitude, any
voice with a decreasing frequency will likewise have a decreasing
amplitude {\cite{meaculpa}}.

Although the waveform's strongest radiation is emitted in voices
corresponding to $l = 2$, $m = 2$, the voices corresponding to $l =
m$, $m \ne 2$ are also quite strong, particularly at late times.  For
example, at the end of inspiral, the $l = 4$, $m = 4$, $k = 0$ voice
is about one fifth as strong as the $l = 2$, $m = 2$, $k = 0$ voice.

\subsection{Waveforms for $a = 0.3594M$}
\label{subsec:waves_0.3594}

I again focus on waves generated during inspiral beginning at $\iota =
20^\circ$ (cf.\ Fig.\ {\ref{fig:traj_0.3594}}).  The phase and
amplitude evolution are qualitatively similar to those shown in Fig.\
{\ref{fig:mvc_0.998_l2_m2}}, so I will not show additional plots.  I
will focus on voices for $l = 2$, $m = 2$, and $k \in [0, 1, 2, 3,
4]$.

As when $a = 0.998M$, the voice corresponding to $k = 0$ is strongest.
Unlike that case, no other voices become very strong towards the end
of inspiral --- when the inspiral ends, the $k = 1$ voice is about a
factor of $10$ weaker than the $k = 0$ voice.  For this spin, the code
was able to reliably compute all voices except the one corresponding
to $k = 4$.  This is because the frequencies are quite a bit lower in
this case, and $|dr^*/dr|\alt 1.7$ over the inspiral domain.  The grid
spacing $\delta r = 0.1M$ works well, except for large harmonic
indices.

Two important points should be noted from the results for other
voices.  Because $\Omega_\theta$ grows monotonically, no voices fade
away, in contrast to the $l = 2$, $m = 0$ voices for $a = 0.998M$.
Also, the voices for $l > 2$ do not become as strong at late times as
they do when $a = 0.998M$.  At late times, the $l = 4$, $m = 4$, $k =
0$ voice is about one twentieth the strength of the $l = 2$, $m = 2$,
$k = 0$ voice.

\section{Conclusion: implications for LISA sources}
\label{sec:conclusion}

The results presented here give the first strong-field gravitational
waveforms for rigorously computed extreme mass ratio inspiral
trajectories.  Because these trajectories are restricted to zero
eccentricity, they do not correspond to inspirals that LISA is likely
to observe.  Nonetheless, they contain interesting features that are
very likely to be present in the general case, and are certainly
worthy of further study.

One of the most interesting features of these extreme mass ratio
inspiral waves is their ``multi-voice'' structure: each $(l,m,k)$
harmonic follows its own phase and amplitude evolution.  The total
waveform is given by the sum of the various voices as the inspiral
progresses.  Provided inspiral is adiabatic, this multi-voice
structure should describe generic inspiral waves as well, with a third
index describing harmonics of the radial frequency $\Omega_r$.

It should be possible to take advantage of the multi-voice structure
of extreme mass ratio inspirals when developing strategies for
analyzing the LISA datastream.  Searches of data from ground-based
detectors such as LIGO will rely in many cases on matched filtering, a
technique that cross-correlates the instrumental data with templates
of a source's gravitational waveform.  Matched filtering works
particularly well when the template accurately models the source's
phase evolution; accurate models for the amplitude are far less
important.  In the case of extreme mass ratio inspiral, it may be far
easier to search for each voice with its simple phase evolution,
$h_{lmk}(t) = {\cal H}_{lmk}(t)e^{i\Psi_{lmk}(t)}$, than to search for
the entire ``chorus'', $h(t) = \sum h_{lmk}(t)$, with its
comparatively complicated phase evolution.  Some signals might only be
detectable in their higher harmonics --- one can imagine using the $m
= 4$ voices to find inspiral into holes so massive that the $m = 2$
harmonics are at frequencies too low to be seen.  Techniques that look
for extreme mass ratio inspirals on a voice-by-voice basis should have
no problem finding those inspirals, provided these high-harmonic
voices are not too weak.  The Fast Chirp Transform {\cite{ricktom}}
may be a computationally effective means of implementing such a
voice-by-voice search.

In this vein, it is interesting to note that the phase evolution of
each voice is dominated by that voice's integrated frequency.  Recall,
from Eqs.\ (\ref{eq:waveform}) and (\ref{eq:multivoicewave}) --
(\ref{eq:Psi_def}) that the evolving phase consists of two pieces
(ignoring the constant offset $m\varphi$, which arises from initial
conditions).  One piece amounts to an integral of that voice's
frequency, $\omega_{mk}$, and contributes $\sim 10^5 - 10^7$ radians
to the inspiral.  The other corresponds to the phase of the complex
amplitude $Z^H_{lmk}$ at a given moment; it only accumulates a few to
a few tens of radians.  Accurately calculating the inspiral
trajectory, and hence having good information about the evolution of
the orbital frequencies, is likely to be more important than knowing
$\Phi_{lmk}$.

The influence of tidal coupling between the inspiraling body and the
black hole is very robust and interesting --- such coupling will
strongly influence the inspiral.  If nature provides black holes that
rotate sufficiently fast, the prolonging of inspiral due to this
coupling will easily be seen in the gravitational-wave data.  This is
a clean, beautiful probe of the strong-field black hole spacetime.

The analysis presented here suggests several possible directions for
future work:

\begin{itemize}

\item Equatorial orbits.  Glampedakis and Kennefick {\cite{kostas_dan}}
are currently finishing an analysis of radiation reaction on
eccentric, equatorial orbits of Kerr black holes; it is an analysis
equivalent to that done in Ref.\ {\cite{paperI}}.  The radiation
reaction data they develop could be used to duplicate this analysis
but to find inspiral sequences for eccentric, equatorial orbits and to
develop the associated inspiral waveforms.  It would be quite valuable
to have results that include the effect of eccentricity and
modulations induced by the radial frequency $\Omega_r$.

\item Generic orbits.  We can in principle calculate gravitational
waves from generic orbits of Kerr black holes --- all we need to do is
specify the source term for a particular orbit in the Teukolsky
equation (\ref{eq:teuk}) and solve.  We are hampered, though by our
inability to compute backreaction on such orbits.  Valuable insight
may be developed by imposing physically motivated constraints to
produce an approximate inspiral trajectory.  The trajectories
presented here indicate that the change in inclination angle is very
small.  Using the Teukolsky equation to compute $\dot E$ and $\dot
L_z$ from the gravitational-wave flux and imposing $\dot\iota = 0$ is
sufficient to fully compute the inspiral trajectory through the
generic orbit parameter space $(r,e,\iota)$.  The gravitational waves
developed in such a study would be valuable tools for exploring the
influence of all three orbital frequencies
$(\Omega_\phi,\Omega_\theta,\Omega_r)$ on the waveform, even if the
waves so produced do not correspond exactly to those in nature.

\item Improved waveforms.  As discussed in Sec.\ {\ref{subsec:grid}},
the parameter space discretization used here was rather crude.  It was
in fact insufficient to accurately compute the amplitudes and phases
of high frequency contributions to the waveforms shown here.  As work
begins to develop data analysis tools for LISA, this analysis should
be refined so that simple limitations such as this discretization are
unimportant.  Also, the parameter space covered in this analysis is
rather inadequate for studying LISA measurements.  The range covered
here was selected simply for computational convenience.  A better
analysis would chose the parameter space covering to span a range of
inspirals corresponding to the masses and observation times that LISA
is likely to measure.

\end{itemize}

Implementing the above items will put on us well on the way to
understanding the data analysis problem with LISA.  Having a broader
set of accurate waveforms will serve as a very useful testbed for
developing analysis tools and for getting a better understanding of
how measurable extreme mass ratio signals are likely to be.

\acknowledgements

I am indebted to Daniel Kennefick for many useful conversations and
advice, particularly during the writing of the numerical code used in
this analysis.  I also thank Lee Lindblom for asking a question which
led me to begin investigating the inspiral trajectories, Sam Finn for
suggesting that the multi-voice decomposition shown in Eq.\
(\ref{eq:multivoicewave}) should be analyzed, and Amy Hughes for
helping develop some of the software used to process the numerical
data shown here.  I thank Lior Burko, Curt Cutler, Yuri Levin, Sterl
Phinney, and Kip Thorne for many useful discussions.  All of the
numerical code used in this analysis was developed with tools from the
Free Software Foundation; all plots were generated with the package
{\sc sm}.  This research was supported at the ITP by NSF Grant
PHY-9907949 and at Caltech by NSF Grants AST-9731698 and AST-9618537,
and NASA Grants NAG5-6840 and NAG5-7034.

\eject


%

\end{document}